\documentclass[apj]{emulateapj}
\usepackage{apjfonts}
\usepackage{hhline}
\usepackage{amsmath}
\usepackage{graphicx}
\usepackage{natbib}
\usepackage{epstopdf} 
\usepackage{subfigure}
\usepackage{color}
\usepackage{comment}
\usepackage{longtable}

 \font\sevenrm=cmr7 scaled 1000

\newcommand{\CIV}{C {\sevenrm IV}}

\newcommand{\HI}{HI}

\begin{document}
\title{A systematic search for X-ray cavities in galaxy clusters, groups, and elliptical galaxies}
\author{Jaejin Shin$^{1}$}
\author{Jong-Hak Woo$^{1,\dag}$}
\author{John S. Mulchaey$^{2}$}

\affil{
$^1$Astronomy Program, Department of Physics and Astronomy, 
Seoul National University, Seoul, 151-742, Republic of Korea\\
$^2$Carnegie Observatories, 813 Santa Barbara St., Pasadena, CA, 91101, USA\\
}

\altaffiltext{\dag}{Author to whom any correspondence should be addressed.}

\begin{abstract}
We perform a comprehensive study of X-ray cavities 
using a large sample of X-ray targets selected from the Chandra archive.  
The sample is selected to cover a large dynamic range 
including galaxy clusters, groups, and individual galaxies. 
Using $\beta$-modeling and unsharp masking techniques, we investigate the presence of 
X-ray cavities for 133 targets that have sufficient X-ray photons for analysis.
We detect 148 X-ray cavities from 69 targets and measure their properties, 
including cavity size, angle, and distance from the center of the diffuse X-ray gas.
We confirm the strong correlation between cavity size and distance from the X-ray 
center similar to previous studies \citep[i.e.,][]{Birzan2004, Diehl2008c, Dong2010}. 
We find that the detection rates of X-ray cavities are similar among galaxy clusters, groups and individual
galaxies, suggesting that the formation mechanism of X-ray cavities is independent of environment.\\

\end{abstract}

\section{INTRODUCTION} \label{section:intro}

The empirical scaling relations between supermassive black holes (SMBHs) 
and their host galaxy properties imply the coevolution of SMBHs and galaxies
\citep[e.g.,][]{Magorrian1998,Ferrarese2000,Gebhardt2000,Kormendy2013,Woo2013},
while understanding the nature of the coevolution remains as one of the 
important open questions.
Representing the mass-accreting phase of SMBHs, active galactic nuclei (AGNs)
have been considered as playing a crucial role through feedback in galaxy
evolution. AGN activities may quench or enhance star formation as several simulations demonstrated both negative feedback 
\citep{Kauffmann2000,Granato2004,DiMatteo2005,
Springel2005,Bower2006,Croton2006,Hopkins2006,Ciotti2010,Scannapieco2012},
and positive feedback \citep{Silk2005,Gaibler2012,Zubovas2013,Ishibashi2013}.

Various observational studies have revealed evidence for AGN feedback 
\citep[][and references therein]{Fabian2012,Heckman2014}, which occurs
mainly in two modes, radiative (quasar) mode and kinetic (radio) mode. 
The radiative mode is related to the radiation pressure from the 
accretion disk, while the kinetic mode is caused by the mechanical 
power of radio jets.

The radiative mode is manifested in many different ways.
Ionized gas outflows are typically observed in type 1 and type 2 AGNs
\citep[e.g.,][]{Crenshaw2003,Nesvadba2008,Nesvadba2011, Woo2016}. 
Winds are detected very close to the central region based on the 
kinematic features of narrow absorption lines in X-ray \citep{Pounds2003,Tombesi2010} 
and broad absorption lines in UV \citep{Weymann1991,Crenshaw2003,Ganguly2007}. 
The asymmetric and blueshifted line profiles of \CIV\ have 
been suggested as evidence for an AGN wind \citep[i.e.,][]{Sulentic2000,Wang2011}.  
At larger scales, velocity dispersions and shifts of emission lines in both ionized and molecular gas
suggest AGN winds on scales comparable to that of the host galaxy
\citep{Boroson2005,Komossa2008,Barbosa2009,Bae2014,Woo2016}.

In contrast, the radio mode feedback has been investigated in more
  massive systems at much larger scales. The radio mode feedback has been
  invoked to reconcile the cooling flow problem, i.e. that the
  observed cooling rate of hot gas in the central regions of galaxies
  and galaxy clusters is lower than the predicted value from cooling
  flow models (see \citealt{Fabian1994}). A natural explanation is the
  existence of a heating source that prevents the cooling of the hot
  gas.  AGNs have been considered as one possible candidate for
  heating (see \citealt{McNamara2007,Fabian2012,Gitti2012} for
  review).  Several studies based on simulations showed that AGN jets
  can provide sufficient energy to affect galaxy evolution and resolve
  the cooling flow problem
  \citep[e.g.,][]{Sijacki2006,Dubois2010,Gaspari2012}.

Based on the high-resolution X-ray imaging obtained with the Chandra
and XMM-Newton telescopes, surface brightness depressions of diffuse
X-ray emission have been detected at the centers of a number of
massive galaxies and galaxy clusters, e.g., Abell 426
(\citealt{Fabian2000,Fabian2006}), Hydra A
\citep{McNamara2000,Nulsen2005a,Wise2007}, M87
\citep{Churazov2001,Forman2007}, Abell 2052 \citep{Blanton2011},
MS0735+7421 \citep{McNamara2005}, Abell 2199 \citep{Johnstone2002},
and Centaurus Cluster \citep{Sanders2002}.  These so-called X-ray
cavities extend up to several hundred kpc (i.e., MS 0735.6+7421),
exceeding the optical size of their host galaxies. Such cavities are
considered a signpost of radio mode feedback, as they are often filled
with radio lobes, implying that the cavities originate from the
interaction between radio jets and interstellar/intergalactic medium
\citep[e.g.,][]{McNamara2000,Fabian2002,McNamara2005,Giacintucci2011,Chon2012}.
In general, observational studies found that the estimated jet energy
inferred from the X-ray cavities can balance the cooling of hot gas
\citep[e.g.,][]{Birzan2004,Rafferty2006,Nulsen2009,Hlavacek2012}.
Thus, studying the physical properties of X-ray cavities can shed
light on understanding AGN feedback and galaxy evolution.

To understand the nature of X-ray cavities, a number of studies have
investigated their properties using samples of galaxies, groups, and
clusters.  For example, by detecting X-ray cavities from 14 galaxy
clusters, one galaxy group, and one galaxy, \cite{Birzan2004}
presented the relationship between X-ray cavity size and the distance
from X-ray center. \cite{Diehl2008c} expanded the sample size to 32
objects (30 galaxy clusters, one galaxy group, and one galaxy) and
found a similar trend between cavity size and distance from the
center.  \cite{Dunn2006} suggested that the AGN feedback duty cycle is
$\sim$ 70\% based on the X-ray cavities detected from 14 galaxy
clusters among a sample of 20 cool-core clusters.  There have also
been statistical studies for smaller scale systems such as galaxy
groups and elliptical galaxies.  For example, \cite{Dong2010} detected
X-ray cavities in 26 galaxy groups, and showed that these systems
follow the same size-distance relation found in galaxy clusters.
X-ray cavities were also detected in giant elliptical galaxies
\citep[e.g., 24 objects out of 104 galaxies,][]{Nulsen2009}.  Based on
the detection rates of X-ray cavities in various environments, 
\cite{Dong2010} discussed the dependence of AGN feedback duty cycle on
environments (i.e., galaxy clusters, galaxy groups, and isolated
elliptical galaxies)

Recent work increased the number of known objects with X-ray cavities.
\cite{Panagoulia2014} identified X-ray cavities from 30 out of 49
local galaxy groups and clusters, which have a cool-core at the
central region. On the other hand, \cite{Hlavacek2012} detected X-ray
cavities in 20 objects, using 76 massive clusters at higher redshifts
($0.3<z<0.7$).  Also, \cite{Hlavacek2015} detected X-ray cavities from
6 objects among 83 galaxy clusters selected with the South Pole
Telescope ($0.4<z<1.2$). Using the size of the cavities,
\cite{Hlavacek2012,Hlavacek2015} claimed that there is no evolution in
radio mode feedback in at least the last $\sim$ 7 Gyr.

While a number of previous studies reported X-ray cavity detections
using various samples, it is difficult to statistically investigate
the physical properties of X-ray cavities as a population due to the
incompleteness and various selection functions of individual studies.
For example, the sample size of
\cite{Dong2010,Hlavacek2012,Panagoulia2014} studies is relatively
large (i.e., 51, 76, and 49 objects, respectively), reporting
discovery of a number of new X-ray cavities.  However, a large
fraction of the targets in these surveys do not have sufficient photon
counts to properly study the presence of X-ray cavities.

In this paper, we present the results from our X-ray cavity study
based on a large sample of targets, for which diffuse X-ray emission
is detected in the $Chandra$ X-ray images. Our sample also covers a
large dynamic range from isolated galaxies to galaxy clusters.  By
adopting a consistent X-ray photon number criterion, we analyze X-ray
images in order to detect X-ray cavities, measure their properties,
and investigate the dependence of the cavity formation on
environments.  We describe the sample selection and data reduction in
\S 2, and the detection method and analysis in \S 3. The main results
are presented in \S 4, followed by discussion in \S 5, and summary and
conclusions in \S 6.  We adopt a cosmology of $H_{\rm 0}= 70$ km
s$^{-1}$ Mpc$^{-1}$, $\Omega_{\Lambda}=0.7$ and $\Omega_{\rm
  m}=0.3$. \\

\section{Sample and data}\label{section:Sample}

To search X-ray cavities, we used available $Chandra$ X-ray images,
which provide high spatial resolution (PSF FWHM $\sim$0\farcs 5 at the
aim point), ideal for X-ray cavity studies.  We considered sources in
the $Chandra$ archive in one of three categories, namely, `normal
galaxies', `cluster of galaxies', and `active galaxies and
quasar'. Using these categories enables us to cover a large dynamical
range.  We selected a parent sample of $\sim$3500 targets, for which
5472 individual exposures are available, including multiple images per
target (i.e., 2668, 1071, and 1733 exposures of active galaxies and
quasars, normal galaxies, and cluster of galaxies respectively).  For
our cavity study, we used a final sample of 133 targets based on a
consistent analysis of X-ray photon counts as described below.

To detect surface brightness depression, a sufficient number of X-ray photons is required. 
In a previous study, \cite{Dong2010} discussed the conditions of X-ray cavity detection based on simulations 
with various parameters (e.g., radial profile $\beta$, total photon count, cavity strength, and distance from the center), 
concluding that total photon count is the most important parameter for X-ray cavity detection. 
However, we found that the total photon count is not a reliable criterion for sample selection since
it strongly depends on the spatial distribution of the diffuse emission.
For example, if the diffuse X-ray photons are spread out with an average small photon count per pixel, it is difficult
to detect cavities although the total photon count is large. On the other hand, if the photon count is high 
at the central part, then cavities can be detected although the total photon count is small.
Thus, we concluded that the total photon count does not provide a consistent selection condition. 

Instead of the total photon count, we focused on the photon counts in
the diffuse gas near the center of the image.  Many targets show an
X-ray AGN at the center, hence, the count from the point source
contributes to the total count, contaminating the diffuse gas
emission. To avoid this contamination, we decided to use an annulus to
measure the photon counts for selecting targets of further analysis.
Since the center of the diffuse emission is typically close to the aim
point of ACIS, we used the size of the PSF FWHM $\sim$2 pixel (i.e.,
0.984 \arcsec), which is relevant for most of energy
bands\footnote{
  http://cxc.harvard.edu/cal/Acis/Cal\_prods/psf/fwhm.html}.  While
for some targets, the diffuse emission is off from the aim point and
the PSF size varies as a function of distance from the aim point, we
assumed PSF size is 2 pixels for all images since we performed PSF
deconvolution as described later in this section.

The contribution of the point source decreases outwards, becoming
$\sim$1\% of the peak value at the distance of 3 pixels in the Chandra PSF.  
We conservatively adopted an annulus with an inner radius of 6 and the outer radius of 7
pixels to calculate the mean count per pixel within the annulus.
After that, we selected targets if the mean X-ray photon count of the
annulus is larger than 4 counts per pixel, which corresponds to
Poisson noise of 2.

Since the sample size is very large, we initially selected targets
with enough X-ray photons by examining the 4$\times$4 pixel binned
images, which was intended to increase the S/N ratio.  We excluded
targets if diffuse X-ray emission is not visible, for which the photon
count corresponds to 2-3 counts per pixel (i.e., 0.1-0.2 counts per
unbinned pixel). In other words, these excluded images obviously have
a far lower photon count that 4 counts per pixel. Thus, this initial
selection conservatively reduced target sample size to $\sim$800
objects.  Note that although we checked the individual images in this
process, we also checked that if multiple exposures are available, the
combined images of these excluded targets do not have sufficient
photon counts.

For the selected $\sim$800 targets, we performed more detailed
analysis after PSF deconvolution.  Since the PSF of Chandra images
varies along with the spatial location from $\sim$0\farcs5 to a few
arcseconds, relatively small cavities far from the center can be
affected by the effect of the PSF.  To deconvolve the PSF of each
image, we reduced raw data using the "Chandra\_repro" script of the
Chandra Interactive Analysis of Observations software (CIAO) v4.6 and
Chandra Calibration Database (CALDB) 4.6.3.  This script performs all
reduction processes such as "acis\_process\_event", including
charge transfer inefficiency correction, time-dependent gain
adjustment, and screening for bad pixels using the bad pixel map in
the pipeline. After reduction process, we generated count images 
and exposure-corrected images in
the 0.5-7 keV energy band, using "$\rm fluximage$" script as well as
the PSF map using "mkpsfmap" script.  Based on the count image
and the PSF map, we performed the PSF deconvolution, using
"deconvblind" function of MATLAB. The majority of the targets in the
sample are observed at the aim point, hence the PSF effect is
negligible for cavity detection. In contrast, if diffuse emission is
far from the aim point, the PSF- deconvolved images significantly
improve the detection of X-ray cavities.
Once the PSF deconvolution is complete, then we combined individual
exposures to improve S/N whenever multiple images are available per
target. 

After PSF deconvolution and combining exposures, we measured the
photon counts at the annulus (6-7) and selected objects with more than
4 counts. With these criteria, we finalized a sample of 133 targets,
which is composed of isolated elliptical galaxies, galaxy groups, and
galaxy clusters, covering a range of gas temperatures and various
environments (i.e., from 0.3 to 10 keV, which are typical temperature
of an isolated elliptical galaxy and a massive galaxy cluster,
respectively).  This sample is the largest to date for investigating
X-ray cavities over a broad dynamic range, enabling a comprehensive
study of X-ray cavities and their formation and evolution in different
environments. The details of our targets are presented in Table 1.
Note that we use exposure-corrected images for cavity analysis
while we use count images for sample selection.

In the sample selection, we have excluded several objects although these objects
satisfied our photon count criteria. First, we excluded spiral
galaxies (i.e., M31) and starburst galaxies (i.e., NGC253, M82), since
diffuse X-ray emission for those target is due to evolved star or
supernova \citep[e.g.,][]{Strickland2000,Strickland2004,Bogdan2008}.
Second, we excluded merging clusters, which show strongly disturbed
distributions (e.g., bullet cluster), using the three lists, namely,
"MCC Radio Relic Sample", "MCC Chandra-Planck Merging Cluster Sample", and "Other Proven Dissociative Mergers", 
from the merging cluster collaboration\footnote{http://www.mergingclustercollaboration.org}. 
Note that we did not exclude two objects (Abell 2146 and Cygnus A) in "Other merging clusters" since they are not classified as dissociative.

Diffuse X-ray emission mainly originates from hot gas, although other
sources may contribute, i.e., cataclysmic variables and coronally
active binary stars.  The X-ray luminosity of these additional sources
is less than $10^{34}\ \rm erg\ s^{-1}$ \citep[e.g.,][]{Sazonov2006},
corresponding to $\sim 0.2 $ counts in total for 200 ks exposure at
the distance of the closest target (NGC 4552), which is based on the
calculation, using the Chandra X-Ray Center Portable, Interactive,
Multi-Mission Simulator
software\footnote{http://cxc.harvard.edu/toolkit/pimms.jsp}. Due to
PSF effects, the photon count per pixel will be even smaller. Thus,
contamination from these additional X-ray sources is negligible for
our extragalactic sources.

Since we selected X-ray targets with sufficient photon counts at the
central part of the targets, the sample may be biased to cool-core
clusters, which are known to have high surface brightness
\citep[e.g.,][]{Panagoulia2014}.  We calculated the cooling time of
each target to investigate whether the cooling time is sufficiently
short.  We adopted the method given by Equation 14 of
\cite{Hudson2010}, which calculates cooling time using the density of
ions, electrons, hydrogen, and temperature derived from $\rm
0.004\ R_{500}$ and the cooling function. 
 A spectrum for each target is extracted within $\rm 0.004\ R_{500}$, 
excluding the central region with a 6 pixel radius to remove the AGN contribution, as similarly used in the sample selection.
Here, we used "specextract" script of CIAO with the longest exposure for each target for simplicity while  
$\rm R_{500}$ was estimated using Equation 12 of \cite{Vikhlinin2006}. 
Since the $\rm R_{500}$ and gas temperature are dependent on each other, we estimated them with iterations. 
The detailed method to estimate $\rm R_{500}$ and gas temperature will be presented in \S3.4. 
Then, we fitted the spectra using XSPEC (version 12.8.2) with a single temperature plasma model (MEKAL) and hydrogen gas absorption model (WABS), with gas temperature, normalization factor, and metal abundance as free parameters. We generally treated
galactic absorption as a free parameter, except for a few sources where we adopted galactic
  \HI\ column density from \cite{Dickey1990} 
  since the parameter did not converge to reliable values due
  to the low photon counts. Based on the normalization factor, we calculated electron density and hydrogen density assuming that the ratio of electrons to hydrogen is 1.2 and that hydrogen and ions have the same density. Finally, the cooling function was adopted from \cite{Sutherland1993} and interpolated for our estimated metal abundances and gas densities.
In summary, we found that the cooling time of all targets is not larger than 3 Gyrs, suggesting that in general our targets have cool cores at the centers.\\

\begin{longtable*}{lcrcrcc}
\caption{Target list}\\
\hline

\multicolumn{1}{c}{Object} & RA & \multicolumn{1}{c}{DEC} & \multicolumn{1}{c}{z} 
& \multicolumn{1}{c}{$N_{H}$} & \multicolumn{1}{c}{T} & \multicolumn{1}{c}{Class}\\
 &  &  &  
& \multicolumn{1}{c}{($10^{20} \rm{cm^{-2}}$)} & \multicolumn{1}{c}{(keV)} & \\

\multicolumn{1}{c}{(1)} & \multicolumn{1}{c}{(2)}  & \multicolumn{1}{c}{(3)} &
\multicolumn{1}{c}{(4)} & \multicolumn{1}{c}{(5)} & \multicolumn{1}{c}{(6)} &
\multicolumn{1}{c}{(7)}\\
\hline
\endfirsthead

\multicolumn{7}{c}%
{\tablename\ \thetable\ -- Continued \textit{}} \\
\hline
\multicolumn{1}{c}{Object} & RA & \multicolumn{1}{c}{DEC} & \multicolumn{1}{c}{z} 
& \multicolumn{1}{c}{$N_{H}$} & \multicolumn{1}{c}{T} & \multicolumn{1}{c}{Class}\\
 &  &  &  
& \multicolumn{1}{c}{($10^{20} \rm{cm^{-2}}$)} & \multicolumn{1}{c}{(keV)} & \\

\hline

\endhead
\endfoot
\hline
\multicolumn{7}{l}{\scriptsize{ \hspace{2ex} Note -- Col. (1): Object name. Col. (2): RA. Col. (3): DEC. Col. (4): Redshift from NASA/IPAC }} \\
\multicolumn{7}{l}{\scriptsize{extragalactic database. Col. (5): HI column density. For targets with an asterisk, the quoted values are}} \\
\multicolumn{7}{l}{\scriptsize{from \cite{Dickey1990}. Col. (6): Gas temperature.  Col. (7):  Class of X-ray cavity detection: 1:}}\\
\multicolumn{7}{l}{\scriptsize{X-ray cavities detected from raw images, 2: X-ray cavities detected from beta-model }}\\
\multicolumn{7}{l}{\scriptsize{subtracted images, 3: and X-ray cavities non-detected.}}

\endlastfoot
Abell 426	&	03	:	19	:	44	&	41	:	25	:	19	&$	0.018	$&	13.66		&$	3.61	_{	-0.01	}^{	+	0.01	}$&	1	\\
NGC 1316	&	03	:	22	:	42	&	-37	:	12	:	29	&$	0.006	$&	5.32		&$	0.74	_{	-0.03	}^{	+	0.03	}$&	1	\\
2A 0335+096	&	03	:	38	:	41	&	09	:	58	:	05	&$	0.036	$&	24.38		&$	1.98	_{	-0.01	}^{	+	0.01	}$&	1	\\
Abell 478	&	04	:	13	:	25	&	10	:	27	:	58	&$	0.088	$&	33.76		&$	4.80	_{	-0.06	}^{	+	0.06	}$&	1	\\
MS 0735.6+7421	&	07	:	41	:	50	&	74	:	14	:	53	&$	0.216	$&	2.51		&$	4.15	_{	-0.10	}^{	+	0.10	}$&	1	\\
HYDRA A	&	09	:	18	:	06	&	-12	:	05	:	46	&$	0.055	$&	4.73		&$	3.44	_{	-0.03	}^{	+	0.06	}$&	1	\\
RBS 797	&	09	:	47	:	13	&	76	:	23	:	17	&$	0.354	$&	5.33		&$	5.07	_{	-0.15	}^{	+	0.16	}$&	1	\\
M84	&	12	:	25	:	04	&	12	:	53	:	13	&$	0.003	$&	7.99		&$	0.74	_{	-0.02	}^{	+	0.02	}$&	1	\\
M87	&	12	:	30	:	50	&	12	:	23	:	31	&$	0.004	$&	7.47		&$	1.72	_{	0.00	}^{	+	0.00	}$&	1	\\
NGC 4552	&	12	:	35	:	37	&	12	:	32	:	38	&$	0.001	$&	2.57*		&$	0.63	_{	-0.03	}^{	+	0.03	}$&	1	\\
NGC 4636	&	12	:	42	:	50	&	02	:	41	:	17	&$	0.003	$&	3.67		&$	0.76	_{	0.00	}^{	+	0.00	}$&	1	\\
Centaurus cluster	&	12	:	48	:	49	&	-41	:	18	:	43	&$	0.011	$&	12.36		&$	2.06	_{	0.00	}^{	+	0.00	}$&	1	\\
HCG 62	&	12	:	53	:	06	&	-09	:	12	:	22	&$	0.015	$&	11.86		&$	0.96	_{	0.00	}^{	+	0.00	}$&	1	\\
NGC 5044	&	13	:	15	:	24	&	-16	:	23	:	06	&$	0.009	$&	10.34		&$	0.98	_{	0.00	}^{	+	0.00	}$&	1	\\
Abell 3581	&	14	:	07	:	30	&	-27	:	01	:	05	&$	0.023	$&	6.95		&$	1.63	_{	-0.01	}^{	+	0.01	}$&	1	\\
NGC 5813	&	15	:	01	:	07	&	01	:	41	:	02	&$	0.007	$&	6.37		&$	0.72	_{	0.00	}^{	+	0.00	}$&	1	\\
Abell 2052	&	15	:	16	:	44	&	07	:	01	:	16	&$	0.035	$&	5.40		&$	2.37	_{	-0.01	}^{	+	0.01	}$&	1	\\
3C 320	&	15	:	31	:	25	&	35	:	33	:	40	&$	0.342	$&	3.78		&$	3.27	_{	-0.23	}^{	+	0.25	}$&	1	\\
Cygnus A	&	19	:	59	:	28	&	40	:	44	:	02	&$	0.056	$&	32.44		&$	5.54	_{	-0.08	}^{	+	0.08	}$&	1	\\
PKS 2153-69	&	21	:	57	:	06	&	-69	:	41	:	24	&$	0.028	$&	2.5*		&$	1.98	_{	-0.04	}^{	+	0.04	}$&	1	\\
3C 444	&	22	:	14	:	26	&	-17	:	01	:	37	&$	0.153	$&	2.43		&$	1.90	_{	-0.04	}^{	+	0.04	}$&	1	\\
\hline																											
Abell 85	&	00	:	41	:	42	&	-09	:	20	:	53	&$	0.055	$&	5.29		&$	4.65	_{	-0.07	}^{	+	0.07	}$&	2	\\
ZwCl 0040+2404	&	00	:	43	:	52	&	24	:	24	:	22	&$	0.083	$&	5.51		&$	2.73	_{	-0.07	}^{	+	0.09	}$&	2	\\
ZwCl 0104+0048	&	01	:	06	:	49	&	01	:	03	:	22	&$	0.255	$&	3.13		&$	2.79	_{	-0.10	}^{	+	0.09	}$&	2	\\
NGC 533	&	01	:	25	:	31	&	01	:	45	:	32	&$	0.019	$&	12.41		&$	1.02	_{	-0.01	}^{	+	0.01	}$&	2	\\
Abell 262	&	01	:	52	:	47	&	36	:	09	:	07	&$	0.017	$&	10.37		&$	1.62	_{	-0.01	}^{	+	0.01	}$&	2	\\
MACS J0242.5-2132	&	02	:	42	:	36	&	-21	:	32	:	28	&$	0.314	$&	4.45		&$	4.13	_{	-0.23	}^{	+	0.26	}$&	2	\\
Abell 383	&	02	:	48	:	03	&	-03	:	31	:	41	&$	0.187	$&	2.60		&$	3.77	_{	-0.15	}^{	+	0.15	}$&	2	\\
AWM 7	&	02	:	54	:	28	&	41	:	34	:	44	&$	0.017	$&	10.62		&$	3.79	_{	-0.05	}^{	+	0.05	}$&	2	\\
MACS J0329.6-0211	&	03	:	29	:	40	&	-02	:	11	:	38	&$	0.450	$&	5.11		&$	5.00	_{	-0.36	}^{	+	0.39	}$&	2	\\
NGC 1399	&	03	:	38	:	11	&	-35	:	41	:	53	&$	0.005	$&	48.92		&$	0.31	_{	-0.02	}^{	+	0.03	}$&	2	\\
NGC 1404	&	03	:	38	:	52	&	-35	:	35	:	35	&$	0.006	$&	1.36*		&$	0.78	_{	-0.01	}^{	+	0.01	}$&	2	\\
RXC J0352.9+1941	&	03	:	52	:	59	&	19	:	40	:	59	&$	0.109	$&	14.49		&$	2.00	_{	-0.04	}^{	+	0.04	}$&	2	\\
MACS J0417.5-1154	&	04	:	17	:	35	&	-11	:	54	:	32	&$	0.440	$&	4.08		&$	6.31	_{	-0.40	}^{	+	0.42	}$&	2	\\
RX J0419.6+0225	&	04	:	19	:	34	&	02	:	28	:	23	&$	0.012	$&	16.03		&$	1.32	_{	0.00	}^{	+	0.00	}$&	2	\\
EXO 0423.4-0840	&	04	:	25	:	51	&	-08	:	33	:	40	&$	0.038	$&	17.68		&$	2.05	_{	-0.09	}^{	+	0.08	}$&	2	\\
Abell 496	&	04	:	33	:	38	&	-13	:	15	:	43	&$	0.033	$&	4.58		&$	4.15	_{	-1.72	}^{	+	1.72	}$&	2	\\
RXC J0439.0+0520	&	04	:	39	:	02	&	05	:	20	:	42	&$	0.208	$&	12.54		&$	3.60	_{	-0.22	}^{	+	0.21	}$&	2	\\
MS 0440.5+0204	&	04	:	43	:	10	&	02	:	10	:	19	&$	0.190	$&	15.97		&$	4.36	_{	-0.19	}^{	+	0.25	}$&	2	\\
 MACS J0744.9+3927 	&	07	:	44	:	53	&	39	:	27	:	29	&$	0.698	$&	5.30		&$	10.30	_{	-1.50	}^{	+	2.75	}$&	2	\\
PKS 0745-19 	&	07	:	47	:	32	&	-19	:	17	:	46	&$	0.103	$&	41.49		&$	5.12	_{	-0.04	}^{	+	0.04	}$&	2	\\
ZwCl 0949+5207	&	09	:	52	:	49	&	51	:	53	:	06	&$	0.214	$&	2.21		&$	4.25	_{	-0.09	}^{	+	0.11	}$&	2	\\
ZwCl 1021+0426	&	10	:	23	:	39	&	04	:	11	:	13	&$	0.285	$&	2.72		&$	5.34	_{	-0.16	}^{	+	0.24	}$&	2	\\
RXC J1023.8-2715	&	10	:	23	:	50	&	-27	:	15	:	22	&$	0.253	$&	5.79		&$	5.06	_{	-0.17	}^{	+	0.18	}$&	2	\\
Abell 1068	&	10	:	40	:	44	&	39	:	57	:	11	&$	0.138	$&	1.20		&$	4.39	_{	-0.10	}^{	+	0.10	}$&	2	\\
Abell 1204	&	11	:	13	:	18	&	17	:	36	:	11	&$	0.171	$&	3.70		&$	3.47	_{	-0.11	}^{	+	0.13	}$&	2	\\
NGC 4104	&	12	:	06	:	38	&	28	:	10	:	26	&$	0.028	$&	8.57		&$	1.57	_{	-0.05	}^{	+	0.04	}$&	2	\\
NGC 4472	&	12	:	29	:	47	&	08	:	00	:	14	&$	0.003	$&	9.46		&$	1.00	_{	0.00	}^{	+	0.00	}$&	2	\\
Abell 1689	&	13	:	11	:	30	&	-01	:	20	:	31	&$	0.183	$&	4.20		&$	9.29	_{	-0.34	}^{	+	0.34	}$&	2	\\
RX J1350.3+0940	&	13	:	50	:	22	&	09	:	40	:	12	&$	0.090	$&	2.59		&$	3.07	_{	-0.10	}^{	+	0.11	}$&	2	\\
ZwCl 1358+6245	&	13	:	59	:	51	&	62	:	31	:	05	&$	0.329	$&	3.62		&$	5.20	_{	-0.33	}^{	+	0.49	}$&	2	\\
Abell 1835	&	14	:	01	:	02	&	02	:	52	:	44	&$	0.253	$&	3.33		&$	5.61	_{	-0.11	}^{	+	0.11	}$&	2	\\
MACS J1423.8+2404	&	14	:	23	:	48	&	24	:	04	:	44	&$	0.543	$&	4.68		&$	4.77	_{	-0.19	}^{	+	0.20	}$&	2	\\
Abell 1991	&	14	:	54	:	31	&	18	:	38	:	31	&$	0.059	$&	6.53		&$	1.88	_{	-0.02	}^{	+	0.02	}$&	2	\\
RXC J1504.1-0248	&	15	:	04	:	08	&	-02	:	48	:	18	&$	0.215	$&	11.21		&$	5.18	_{	-0.09	}^{	+	0.09	}$&	2	\\
NGC 5846	&	15	:	06	:	29	&	01	:	36	:	22	&$	0.006	$&	10.39		&$	0.81	_{	-0.01	}^{	+	0.01	}$&	2	\\
MS 1512-cB58  	&	15	:	14	:	22	&	36	:	36	:	25	&$	2.723	$&	2.42		&$	7.43	_{	-1.11	}^{	+	1.09	}$&	2	\\
 RXC J1524.2-3154	&	15	:	24	:	13	&	-31	:	54	:	25	&$	0.103	$&	16.05		&$	2.65	_{	-0.03	}^{	+	0.03	}$&	2	\\
RX J1532.9+3021	&	15	:	32	:	54	&	30	:	21	:	00	&$	0.345	$&	2.52		&$	4.44	_{	-0.12	}^{	+	0.12	}$&	2	\\
Abell 2199	&	16	:	28	:	38	&	39	:	33	:	04	&$	0.030	$&	4.05		&$	3.68	_{	-0.04	}^{	+	0.04	}$&	2	\\
Abell 2204	&	16	:	32	:	47	&	05	:	34	:	34	&$	0.152	$&	10.78		&$	4.89	_{	-0.06	}^{	+	0.06	}$&	2	\\
NGC 6338	&	17	:	15	:	23	&	57	:	24	:	40	&$	0.027	$&	10.56		&$	1.68	_{	-0.02	}^{	+	0.02	}$&	2	\\
MACS J1720.3+3536	&	17	:	20	:	17	&	35	:	36	:	25	&$	0.391	$&	2.77		&$	5.38	_{	-0.43	}^{	+	0.58	}$&	2	\\
MACS J1931.8-2634	&	19	:	31	:	50	&	-26	:	34	:	34	&$	0.352	$&	9.84		&$	4.87	_{	-0.13	}^{	+	0.14	}$&	2	\\
MACS J2046.0-3430	&	20	:	46	:	01	&	-34	:	30	:	18	&$	0.423	$&	5.79		&$	4.65	_{	-0.38	}^{	+	0.41	}$&	2	\\
MACS J2229.7-2755	&	22	:	29	:	45	&	-27	:	55	:	37	&$	0.324	$&	3.65		&$	3.73	_{	-0.23	}^{	+	0.24	}$&	2	\\
Sersic 159-03	&	23	:	13	:	58	&	-42	:	43	:	34	&$	0.058	$&	4.53		&$	2.43	_{	-0.02	}^{	+	0.02	}$&	2	\\
Abell 2597	&	23	:	25	:	20	&	-12	:	07	:	26	&$	0.085	$&	3.27		&$	3.13	_{	-0.03	}^{	+	0.03	}$&	2	\\
Abell 2626	&	23	:	36	:	30	&	21	:	08	:	46	&$	0.055	$&	6.51		&$	2.79	_{	-0.04	}^{	+	0.04	}$&	2	\\
\hline																											
MACS J0159.8-0849	&	01	:	59	:	49	&	-08	:	49	:	59	&$	0.405	$&	5.71		&$	5.99	_{	-0.45	}^{	+	0.48	}$&	3	\\
NGC 1132	&	02	:	52	:	52	&	-01	:	16	:	34	&$	0.023	$&	11.78		&$	1.19	_{	-0.02	}^{	+	0.02	}$&	3	\\
Abell 3112	&	03	:	17	:	58	&	-44	:	14	:	17	&$	0.075	$&	4.66		&$	3.88	_{	-0.06	}^{	+	0.06	}$&	3	\\
NGC 1332	&	03	:	26	:	17	&	-21	:	20	:	10	&$	0.005	$&	18.49		&$	0.50	_{	-0.10	}^{	+	0.06	}$&	3	\\
NGC 1407	&	03	:	40	:	12	&	-18	:	34	:	48	&$	0.006	$&	14.99		&$	0.92	_{	-0.02	}^{	+	0.02	}$&	3	\\
4C+37.11	&	04	:	05	:	49	&	38	:	03	:	32	&$	0.055	$&	76.06		&$	3.66	_{	-0.07	}^{	+	0.07	}$&	3	\\
MACS J0429.6-0253	&	04	:	29	:	36	&	-02	:	53	:	06	&$	0.399	$&	5.90		&$	5.34	_{	-0.43	}^{	+	0.57	}$&	2	\\
RXC J0528.9-3927	&	05	:	28	:	56	&	-39	:	27	:	47	&$	0.284	$&	2.45		&$	7.00	_{	-0.53	}^{	+	0.57	}$&	3	\\
PLCKG266.6-27.3	&	06	:	15	:	52	&	-57	:	46	:	52	&$	0.972	$&	4.33		&$	14.02	_{	-1.57	}^{	+	1.57	}$&	3	\\
4C+55.16	&	08	:	34	:	55	&	55	:	34	:	23	&$	0.241	$&	6.45		&$	3.28	_{	-0.06	}^{	+	0.06	}$&	3	\\
MS 0839.9+2938	&	08	:	42	:	56	&	29	:	27	:	29	&$	0.194	$&	6.46		&$	3.30	_{	-0.13	}^{	+	0.13	}$&	3	\\
ZwCl 0857+2107	&	09	:	00	:	37	&	20	:	53	:	42	&$	0.235	$&	3.88		&$	3.13	_{	-0.09	}^{	+	0.09	}$&	3	\\
IRAS 09104+4109	&	09	:	13	:	46	&	40	:	56	:	28	&$	0.442	$&	2.67		&$	4.62	_{	-0.21	}^{	+	0.22	}$&	3	\\
Abell 907	&	09	:	58	:	22	&	-11	:	03	:	50	&$	0.153	$&	7.93		&$	4.78	_{	-0.16	}^{	+	0.16	}$&	3	\\
NGC 3402	&	10	:	50	:	27	&	-12	:	50	:	28	&$	0.015	$&	12.17		&$	0.93	_{	-0.01	}^{	+	0.01	}$&	3	\\
MACS J1115.8+0129	&	11	:	15	:	52	&	01	:	29	:	53	&$	0.352	$&	4.74		&$	5.37	_{	-0.29	}^{	+	0.58	}$&	3	\\
Abell 1361	&	11	:	43	:	40	&	46	:	21	:	22	&$	0.117	$&	5.24		&$	3.23	_{	-0.13	}^{	+	0.13	}$&	3	\\
Abell 1413	&	11	:	55	:	10	&	23	:	30	:	47	&$	0.143	$&	6.52		&$	6.59	_{	-0.19	}^{	+	0.19	}$&	3	\\
RX J1159.8+5531 	&	11	:	59	:	51	&	55	:	32	:	02	&$	0.081	$&	6.65		&$	1.68	_{	-0.02	}^{	+	0.02	}$&	3	\\
MKW 4	&	12	:	04	:	27	&	01	:	53	:	42	&$	0.020	$&	7.92		&$	1.76	_{	-0.03	}^{	+	0.03	}$&	3	\\
NGC 4261	&	12	:	19	:	23	&	05	:	49	:	30	&$	0.007	$&	14.48		&$	0.73	_{	-0.02	}^{	+	0.02	}$&	3	\\
NGC 4649	&	12	:	43	:	40	&	11	:	33	:	11	&$	0.004	$&	13.12		&$	0.89	_{	0.00	}^{	+	0.00	}$&	3	\\
MRK 231	&	12	:	56	:	14	&	56	:	52	:	26	&$	0.042	$&	1.26*		&$	0.83	_{	-0.02	}^{	+	0.02	}$&	3	\\
Abell 1650	&	12	:	58	:	42	&	-01	:	45	:	43	&$	0.084	$&	3.71		&$	5.45	_{	-0.13	}^{	+	0.18	}$&	3	\\
Abell 1664	&	13	:	03	:	42	&	-24	:	14	:	46	&$	0.128	$&	10.23		&$	2.57	_{	-0.06	}^{	+	0.06	}$&	3	\\
MACS J1311.0-0311  	&	13	:	11	:	02	&	-03	:	10	:	37	&$	0.494	$&	4.60		&$	4.47	_{	-0.39	}^{	+	0.47	}$&	3	\\
RX J1347.5-1145	&	13	:	47	:	31	&	-11	:	45	:	11	&$	0.451	$&	6.28		&$	10.46	_{	-0.45	}^{	+	0.53	}$&	3	\\
Abell 1795	&	13	:	48	:	53	&	26	:	35	:	28	&$	0.062	$&	2.03		&$	4.28	_{	-0.05	}^{	+	0.07	}$&	3	\\
MACS J1427.2+4407  	&	14	:	27	:	16	&	44	:	07	:	30	&$	0.487	$&	3.21		&$	4.58	_{	-0.51	}^{	+	0.52	}$&	3	\\
RCS J1447+0828	&	14	:	47	:	27	&	08	:	28	:	19	&$	0.380	$&	5.46		&$	4.88	_{	-0.28	}^{	+	0.29	}$&	3	\\
MS 1455.0+2232	&	14	:	57	:	14	&	22	:	20	:	38	&$	0.258	$&	5.39		&$	4.25	_{	-0.09	}^{	+	0.10	}$&	3	\\
RXC J1459.4-1811	&	14	:	59	:	29	&	-18	:	10	:	44	&$	0.236	$&	9.89		&$	3.58	_{	-0.11	}^{	+	0.11	}$&	3	\\
Abell 2029	&	15	:	10	:	56	&	05	:	44	:	42	&$	0.077	$&	5.04		&$	6.23	_{	-0.05	}^{	+	0.05	}$&	3	\\
MKW 03S	&	15	:	21	:	52	&	07	:	42	:	32	&$	0.045	$&	5.13		&$	3.39	_{	-0.03	}^{	+	0.03	}$&	3	\\
Abell 2146	&	15	:	56	:	10	&	66	:	21	:	25	&$	0.234	$&	2.81*		&$	4.24	_{	-0.15	}^{	+	0.17	}$&	3	\\
Abell 2142	&	15	:	58	:	20	&	27	:	14	:	46	&$	0.091	$&	6.21		&$	7.19	_{	-0.13	}^{	+	0.13	}$&	3	\\
RXC J1558.3-1410	&	15	:	58	:	22	&	-14	:	09	:	58	&$	0.097	$&	12.28		&$	3.65	_{	-0.08	}^{	+	0.08	}$&	3	\\
ESO 137-006	&	16	:	15	:	04	&	-60	:	54	:	25	&$	0.018	$&	21.30		&$	0.99	_{	-0.15	}^{	+	0.12	}$&	3	\\
MACS J1621.3+3810	&	16	:	21	:	25	&	38	:	10	:	08	&$	0.465	$&	2.74		&$	5.00	_{	-0.57	}^{	+	0.72	}$&	3	\\
Abell 2219	&	16	:	40	:	20	&	46	:	42	:	40	&$	0.226	$&	3.19		&$	11.07	_{	-1.10	}^{	+	1.36	}$&	3	\\
Hercules A	&	16	:	51	:	08	&	04	:	59	:	35	&$	0.155	$&	7.18		&$	3.15	_{	-0.11	}^{	+	0.11	}$&	3	\\
Abell 2244	&	17	:	02	:	43	&	34	:	03	:	36	&$	0.097	$&	3.95		&$	5.16	_{	-0.10	}^{	+	0.10	}$&	3	\\
Ophiuchus cluster	&	17	:	12	:	28	&	-23	:	22	:	12	&$	0.028	$&	38.59		&$	7.40	_{	-0.15	}^{	+	0.15	}$&	3	\\
B3 1715+425	&	17	:	17	:	19	&	42	:	26	:	56	&$	0.183	$&	2.40		&$	4.24	_{	-0.31	}^{	+	0.40	}$&	3	\\
RXC J1720.1+2637	&	17	:	20	:	09	&	26	:	38	:	06	&$	0.164	$&	7.27		&$	4.80	_{	-0.16	}^{	+	0.17	}$&	3	\\
ZwCl 1742+3306	&	17	:	44	:	14	&	32	:	59	:	28	&$	0.076	$&	6.65		&$	2.92	_{	-0.05	}^{	+	0.05	}$&	3	\\
NGC 6482	&	17	:	51	:	49	&	23	:	04	:	19	&$	0.013	$&	13.93		&$	0.75	_{	-0.01	}^{	+	0.01	}$&	3	\\
H 1821+643	&	18	:	21	:	57	&	64	:	20	:	38	&$	0.297	$&	4.04*		&$	4.91	_{	-0.14	}^{	+	0.14	}$&	3	\\
NGC 6861	&	20	:	07	:	19	&	-48	:	22	:	12	&$	0.009	$&	6.95		&$	1.25	_{	-0.03	}^{	+	0.03	}$&	3	\\
PKS 2005-489	&	20	:	09	:	25	&	-48	:	49	:	55	&$	0.071	$&	3.08		&$	2.28	_{	-0.12	}^{	+	0.14	}$&	3	\\
 RXC J2014.8-2430	&	20	:	14	:	52	&	-24	:	30	:	22	&$	0.161	$&	14.91		&$	4.05	_{	-0.09	}^{	+	0.10	}$&	3	\\
RX J2129.6+0005	&	21	:	29	:	40	&	00	:	05	:	20	&$	0.235	$&	3.98		&$	5.29	_{	-0.27	}^{	+	0.37	}$&	3	\\
MS 2137.3-2353	&	21	:	40	:	15	&	-23	:	39	:	40	&$	0.313	$&	4.96		&$	4.34	_{	-0.13	}^{	+	0.16	}$&	3	\\
Abell 2390	&	21	:	53	:	36	&	17	:	41	:	46	&$	0.228	$&	11.31		&$	5.92	_{	-0.15	}^{	+	0.15	}$&	3	\\
3C 438	&	21	:	55	:	52	&	38	:	00	:	29	&$	0.290	$&	29.76		&$	6.73	_{	-0.45	}^{	+	0.56	}$&	3	\\
IC 1459	&	22	:	57	:	11	&	-36	:	27	:	43	&$	0.006	$&	10.33		&$	0.57	_{	-0.16	}^{	+	0.05	}$&	3	\\
Abell 2550	&	23	:	11	:	37	&	-21	:	44	:	28	&$	0.123	$&	2.95		&$	2.00	_{	-0.06	}^{	+	0.06	}$&	3	\\
NGC 7618	&	23	:	19	:	47	&	42	:	51	:	07	&$	0.017	$&	22.33		&$	0.91	_{	-0.01	}^{	+	0.01	}$&	3	\\
NGC 7619	&	23	:	20	:	06	&	08	:	09	:	29	&$	0.013	$&	12.01		&$	0.88	_{	-0.02	}^{	+	0.02	}$&	3	\\
Abell 2589	&	23	:	23	:	57	&	16	:	46	:	41	&$	0.041	$&	4.65		&$	3.81	_{	-0.08	}^{	+	0.08	}$&	3	\\
RCS J2327-0204	&	23	:	27	:	28	&	-02	:	04	:	37	&$	0.700	$&	3.82		&$	8.66	_{	-0.85	}^{	+	1.26	}$&	3	\\
Abell 2627	&	23	:	36	:	42	&	23	:	55	:	05	&$	0.126	$&	4.31*		&$	3.80	_{	-0.39	}^{	+	0.40	}$&	3	\\
SPT-CL J2344-4243 	&	23	:	44	:	42	&	-42	:	42	:	54	&$	0.620	$&	2.62		&$	7.47	_{	-0.73	}^{	+	0.78	}$&	3	\\
Abell 4059	&	23	:	57	:	00	&	-34	:	45	:	29	&$	0.049	$&	3.84		&$	3.22	_{	-0.03	}^{	+	0.03	}$&	3	
\end{longtable*}

\begin{figure*}
\includegraphics[width = 0.96\textwidth]{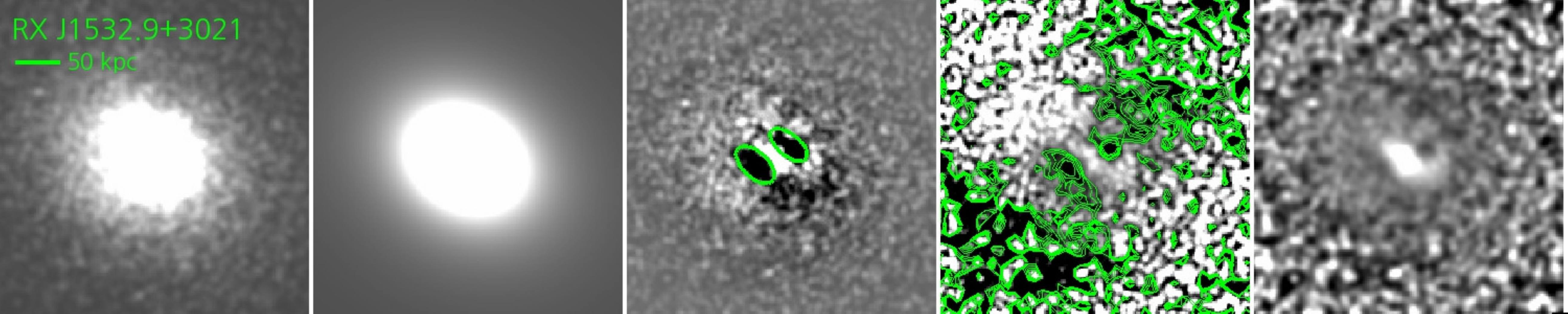}
\caption{
     Example of the $\beta$-model fitting result for RXJ1532.9+3021. Smoothed raw image (left), $\beta$-model (center left), residual image after subtracting
     $\beta$-model (center), normalized image with contour levels of 90\%, 85\%, 80\%, 75\%, 70\% decreasing from outer to inner side (center right) and unsharp masked image (right).
\label{fig:allspec1}}
\end{figure*}

\section{Analysis} \label{section:anlysis}
\subsection{Detection method} \label{section:anlysis}

To detect X-ray cavities, we used two methods, the $\beta$-modeling and the
unsharp masking technique (see Figure 1), as previously used in other studies \citep[e.g.,][]{Diehl2008c,Dong2010}.
First, for $\beta$-modeling, the radial distribution of the surface
brightness of diffuse X-ray gas is modeled assuming an isothermal
distribution of the hot gas \citep{Cavaliere1976}.  Using the $Sherpa$
package in CIAO, we fitted the surface brightness profile with various
parameters, e.g., radius, ellipticity, and power-law index, as free
parameters except for the location of the center of X-ray emission.
We fixed the peak of the distribution as the center since many
  targets show asymmetric distribution (e.g., cold front).  If we fit
  the center as a free parameter, the center is almost identical to
  the peak location within $\sim$1 pixel for the targets that show
  symmetric distribution (e.g., Abell 85, NGC 533, Abell 383), while for the
  targets with asymmetric distribution, the center is significantly
  off from the location of the peak by up to 60 pixels (e.g., Abell
  2146, Ophiuchus cluster). To subtract background,
we used constant value as representative of the background.  We did
not exclude the central bright point source (X-ray AGN) since masking
or fitting the point source does not improve the results.

For the targets with strong and large X-ray cavities (21 objects), the
$\beta$-model does not provide a good fit since the spatial profile
dramatically changes (see Figure 2). Instead, we can directly detect
cavities from the raw images, or use the unsharp masked image to
identify cavities.

Second, we applied the unsharp masking technique by utilizing the
smoothing task "aconvolve" in CIAO.  Adopting 2 pixels (i.e.,
0.984\arcsec) for small scales and 10 pixels for large scales, we
produced smoothed (small and large) images and divided the large scale
smoothed image by the small scale smoothed image to obtain the unsharp
masked image.  A weakness of the unsharp masking technique is that it
is difficult to determine the size of X-ray cavities because the size
varies with smoothing lengths \citep[see][]{Dong2010}.  Also, for
extreme cases, X-ray cavities are not detected from the unsharp masked
image (e.g., NGC 533), while cavities are clearly detected from the
$\beta$-modeling analysis (see Figure 3).  Therefore, we mainly used
the $\beta$-model to detect X-ray cavities, and used the unsharp
masking technique as a secondary method.

 In the detection process, we also used the normalized images which were generated by  dividing
a raw image by the best $\beta$-model, in order to conservatively define cavities more. 
By examining the normalized images as well as residual images, we investigated the spatial extent of relative depression.
We require a minimum of 15\% depression to confirm the presence of cavities and estimate the size for most targets. 
However, if the cavity is clearly present in the raw images, we further classified these features as
X-ray cavities even though the relative depression is slightly lower than 15\%. These exceptions are namely, 
ZwCl 0104+0048, Abell 383, ZwCl 1021+0426, NGC 4104, and MACS J2046.0-3430 (see Figure 3).

In Figure 1, we present an example of the fitting result of RX
J1532.9+3021. Compared to the raw image (left), the residual image (center), after subtracting the
best-fit $\beta$-model (center left) shows two cavities very clearly with the green lines 
representing the ellipse models. The normalized image (center right) confirms the presence of cavities 
 (see Figure 1 for more details). For all 133 targets in the sample, we present the results: directly
detected cavities from raw images (Figure 2), cavities detected via
$\beta$-modeling (Figure 3), and uncertain or non-detections (Figure
4).\\

\begin{figure*}
\includegraphics[width = \textwidth]{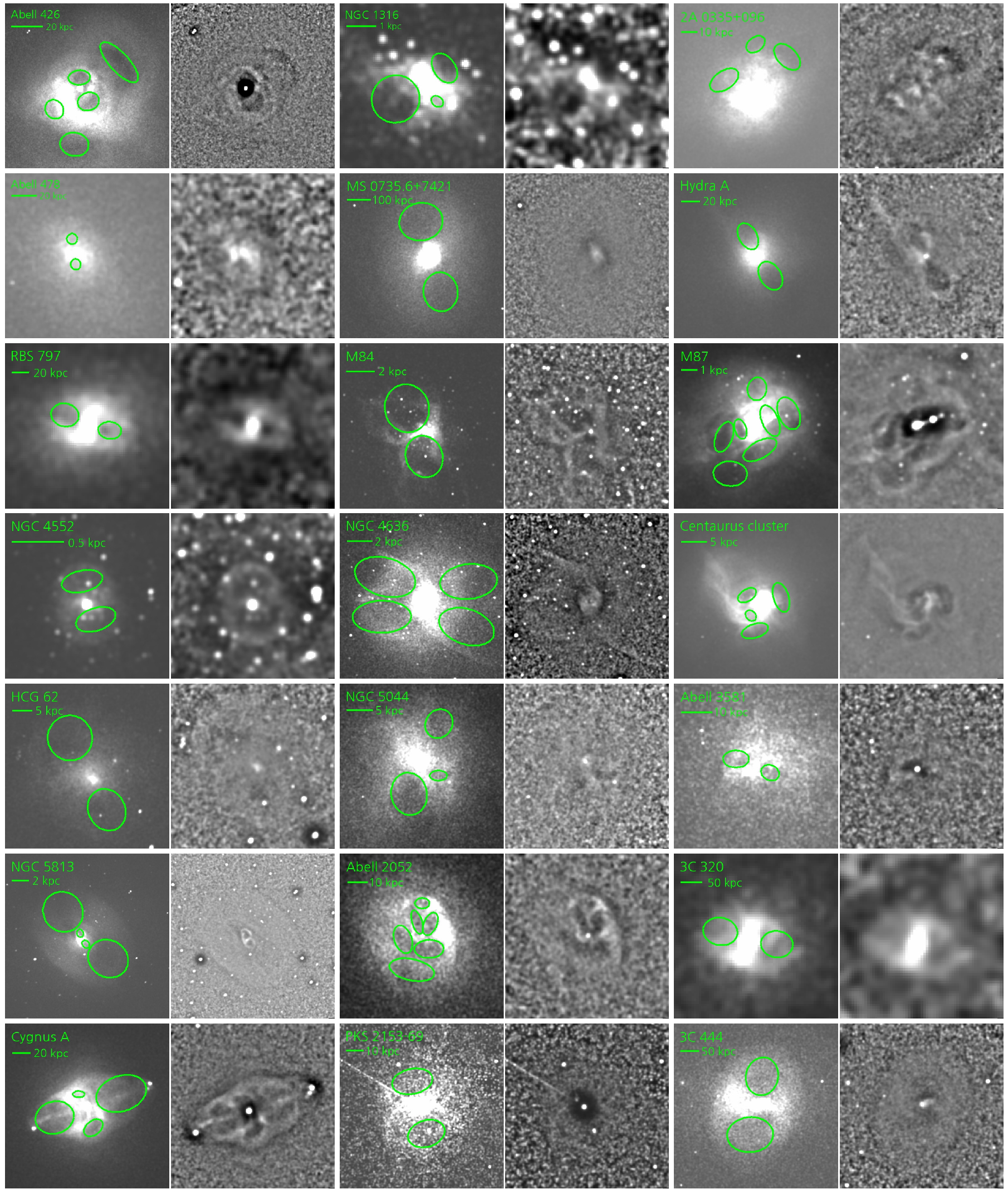}
\caption{
     Targets with cavity detections from raw images . Smoothed raw image (left) and unsharp masked image (right) are shown for each object. 
     Detected cavities are denoted with green ellipse.
\label{fig:allspec1}}
\end{figure*}

\begin{figure*}
\includegraphics[width = \textwidth]{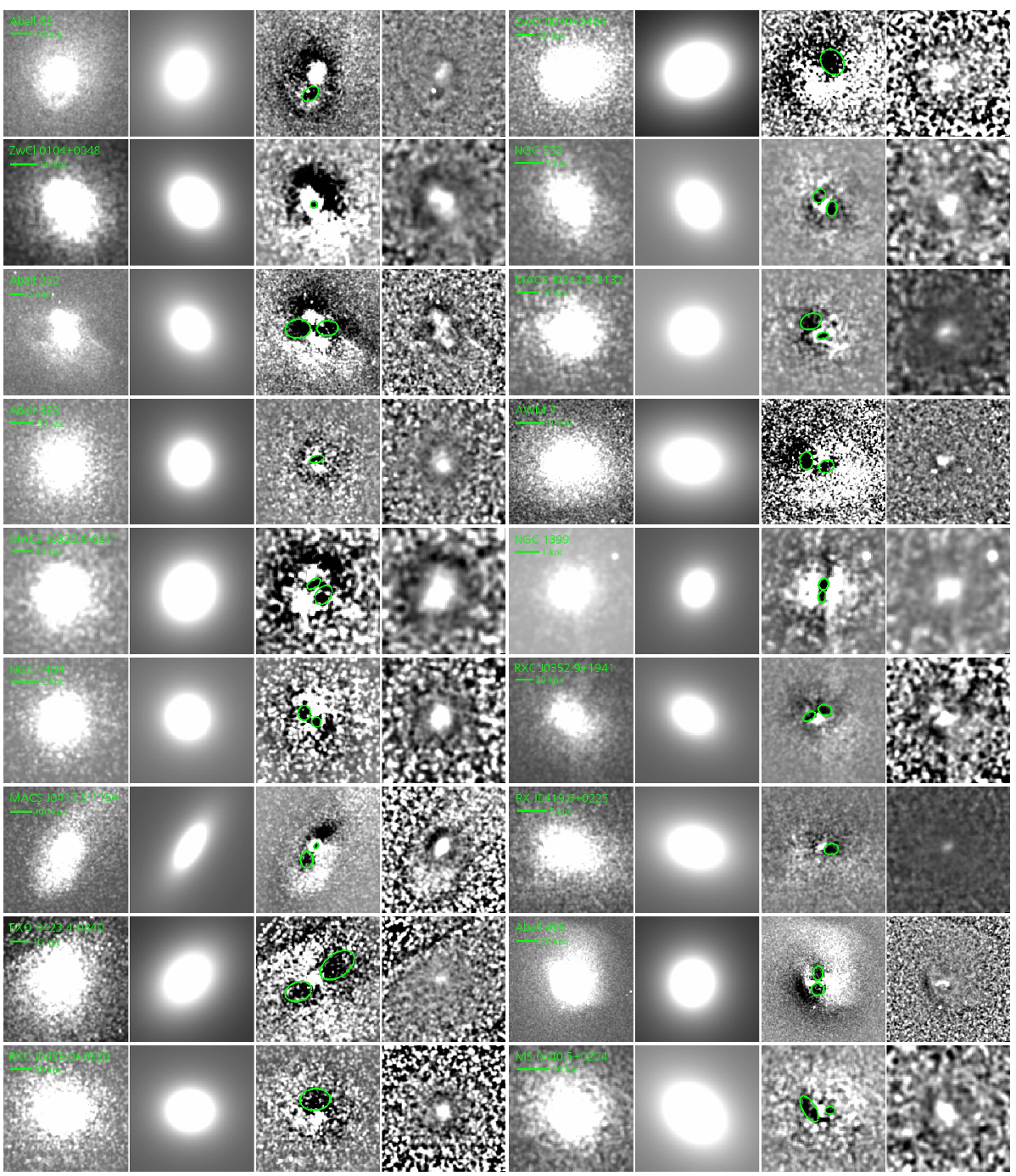} 
\caption{
     Targets with cavity detections from the $\beta$-modeling. From left to right, smoothed raw image, $\beta$-model, residual image, and unsharp masked image are shown. Detected cavities are denoted with green ellipse.}
\label{figure3}
\end{figure*} 

\renewcommand{\thefigure}{\arabic{figure}}
\addtocounter{figure}{-1}
\begin{figure*}
\includegraphics[width = \textwidth]{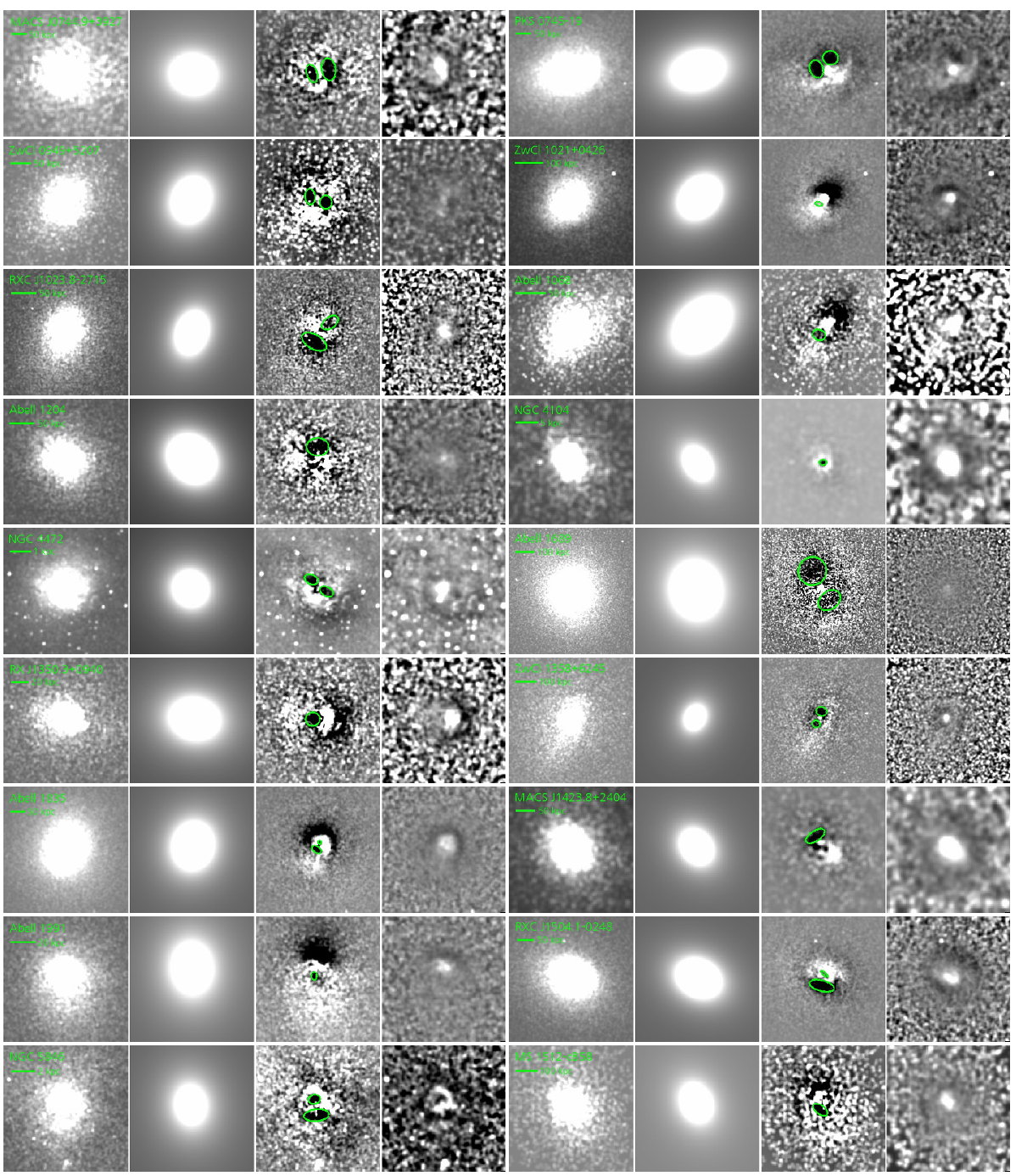}
\caption{ \it continued}
\label{figure5}
\end{figure*} 

\renewcommand{\thefigure}{\arabic{figure}}
\addtocounter{figure}{-1}
\begin{figure*}
\includegraphics[width = \textwidth]{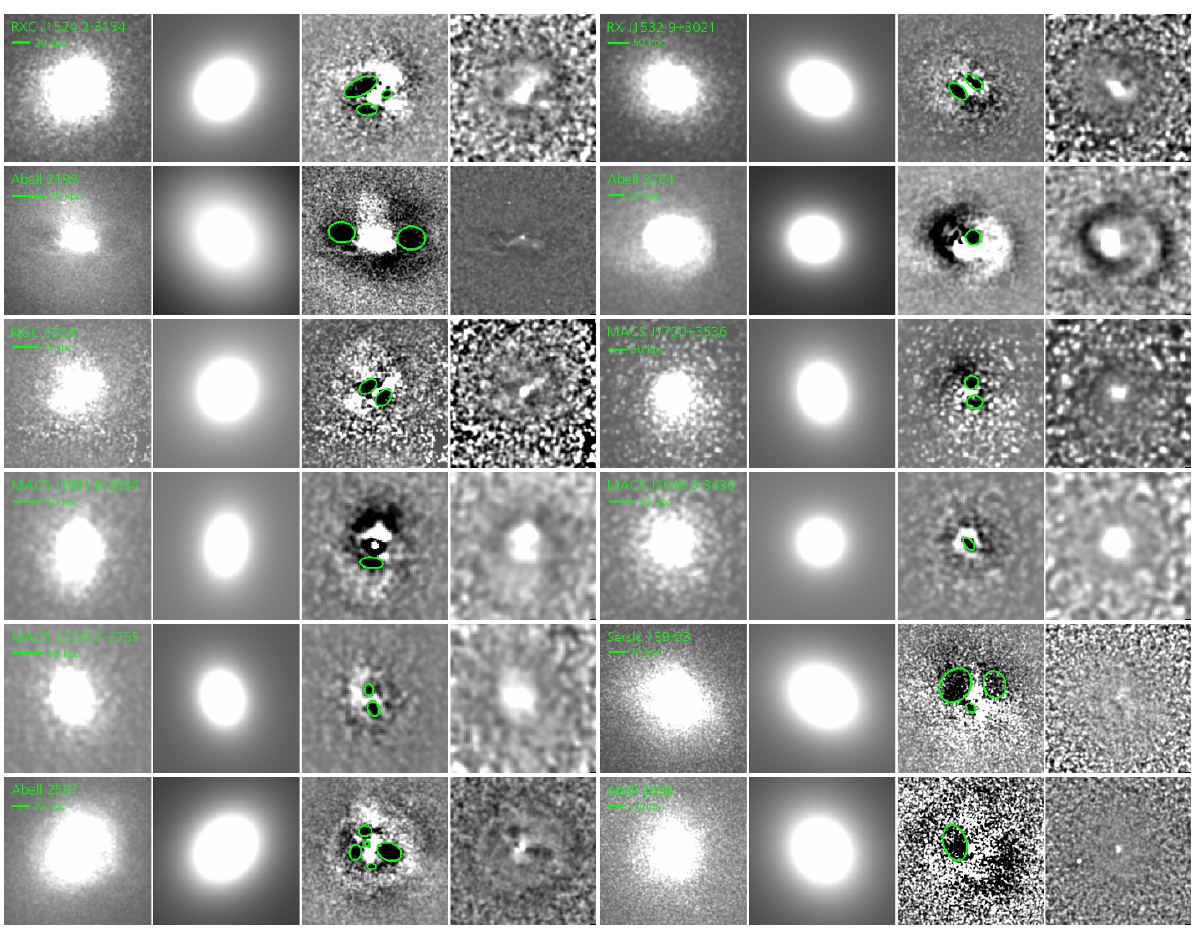}
\caption{ \it continued}
\label{figure5}
\end{figure*} 

\begin{figure*}
\includegraphics[width = \textwidth]{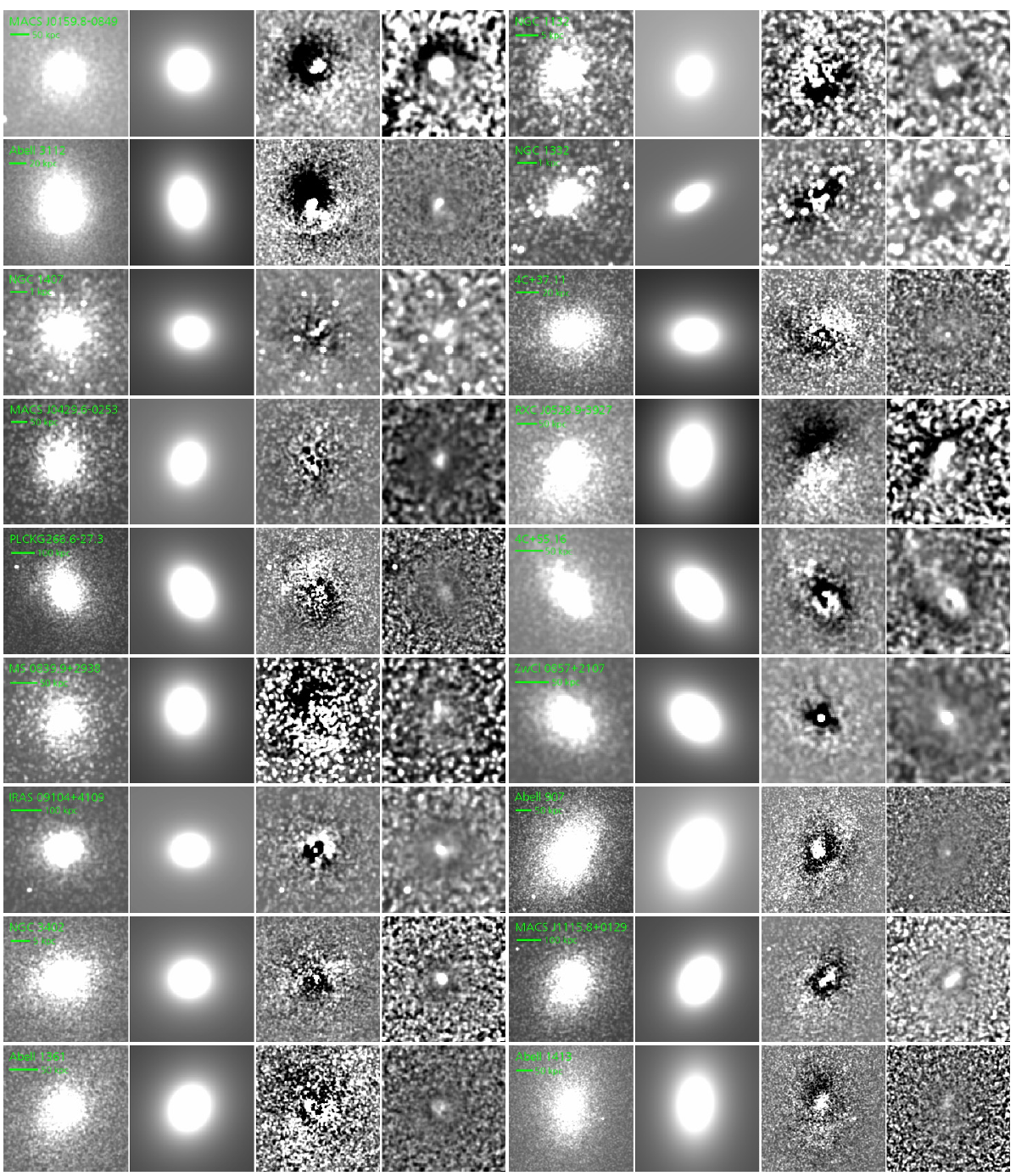}
\caption{
     Targets with non detections.  From left to right, smoothed raw image, $\beta$-model, residual image, and unsharp masked image are shown. 
\label{fig:allspec1}}
\end{figure*} 

\renewcommand{\thefigure}{\arabic{figure}}
\addtocounter{figure}{-1}
\begin{figure*}
\includegraphics[width = \textwidth]{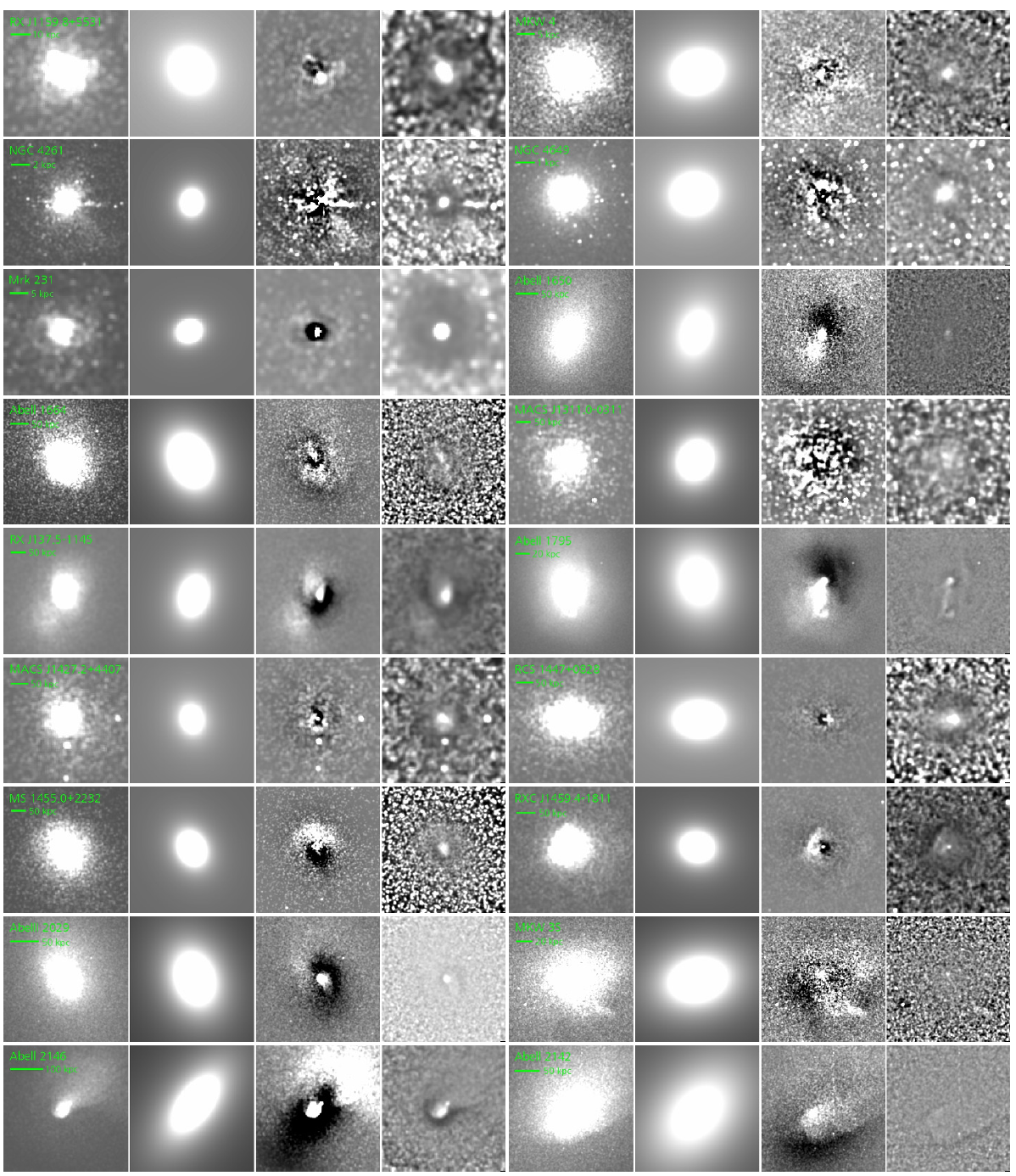}
\caption{ \it continued}
\label{fig:allspec1}
\end{figure*} 

\renewcommand{\thefigure}{\arabic{figure}}
\addtocounter{figure}{-1}
\begin{figure*}
\includegraphics[width = \textwidth]{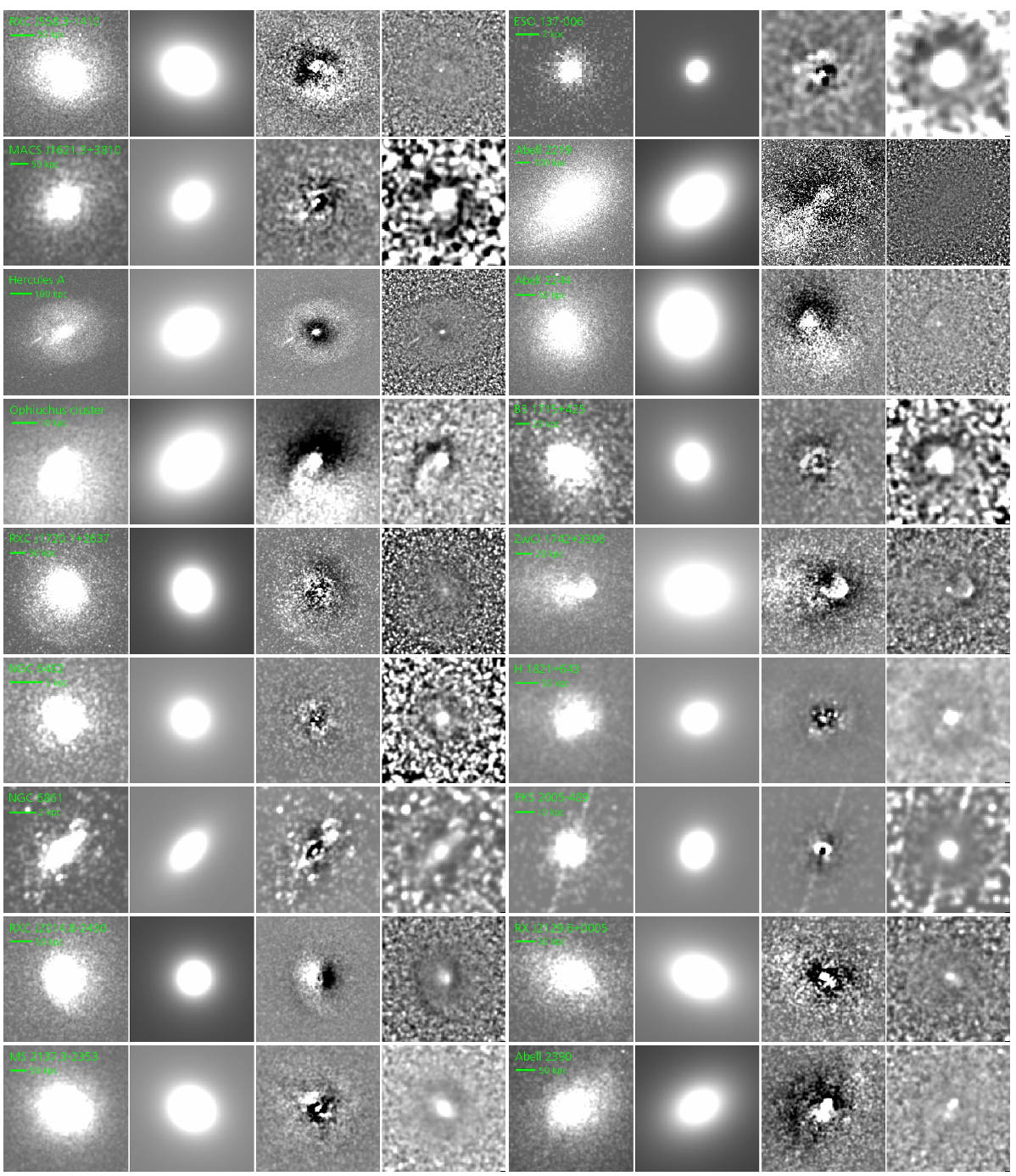}
\caption{ \it continued}
\label{fig:allspec1}
\end{figure*} 

\renewcommand{\thefigure}{\arabic{figure}}
\addtocounter{figure}{-1}
\begin{figure*}
\includegraphics[width = \textwidth]{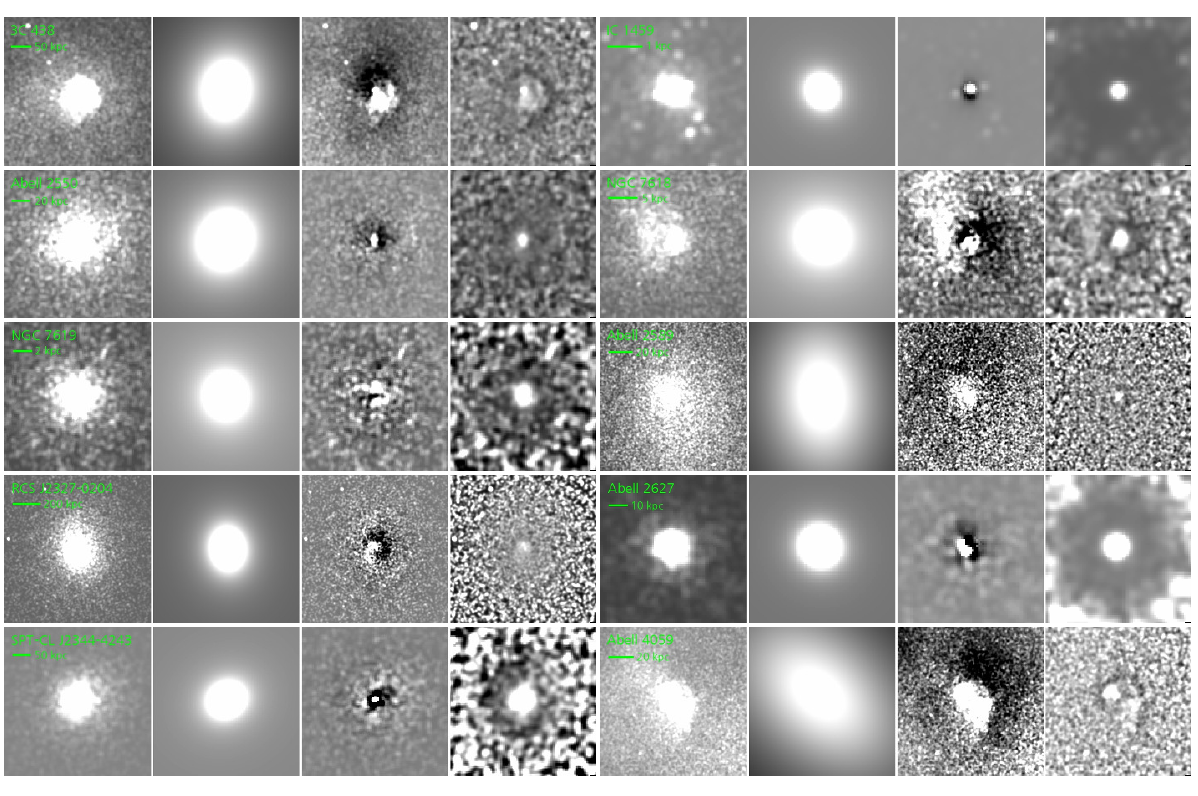}
\caption{ \it continued}
\label{fig:allspec1}
\end{figure*}

\subsection{Cavity properties}

We measured the properties of X-ray cavities, either directly detected
from the raw X-ray images (for 21 targets) or detected from the
$\beta$-model analysis (for 48 targets). Since the shape of X-ray
cavities typically looks elliptical, we adopted an ellipse model to
describe the morphology of X-ray cavities.  Note that we tried to fit
the shape of the individual X-ray cavities, but failed mainly due to
the low photon counts. Thus, we simply overlapped an ellipse model to
the raw and beta-model subtracted images, as similarly performed in
the previous studies \citep{Dong2010}.  Based on the ellipse model, we
determined the size in the semimajor- and semiminor-axes of the X-ray
cavities.  Note that the size of cavities depends on the choice of the
best ellipse model, hence the systematic uncertainty of the size can
be significant (see Section 5.4 for more detailed discussion).

We also measured the distance between the cavity center (i.e., center
of the ellipse) and the center of the diffuse X-ray emission, which is
assumed to be the location of the point source.  If a point source is
not present, we adopted the peak position of diffuse X-ray emission as
the center.  We assumed 2 pixels as the measurement uncertainties of
the cavity size and the distance.  For three targets (ZwCl 0104+0048, NGC 4104, 
and MACS J2046.0-3430), the detected X-ray
cavities are at the center of the diffuse X-ray emission. Perhaps, the
radio jet direction is along the line-of-sight in these cases \citep[see
  also][]{Dong2010}.  In Table 2, we list the properties of X-ray
cavities for individual targets.

\subsection{False detections of X-ray cavities} \label{section: cold front}

\begin{figure*}
\includegraphics[width = 0.95\textwidth]{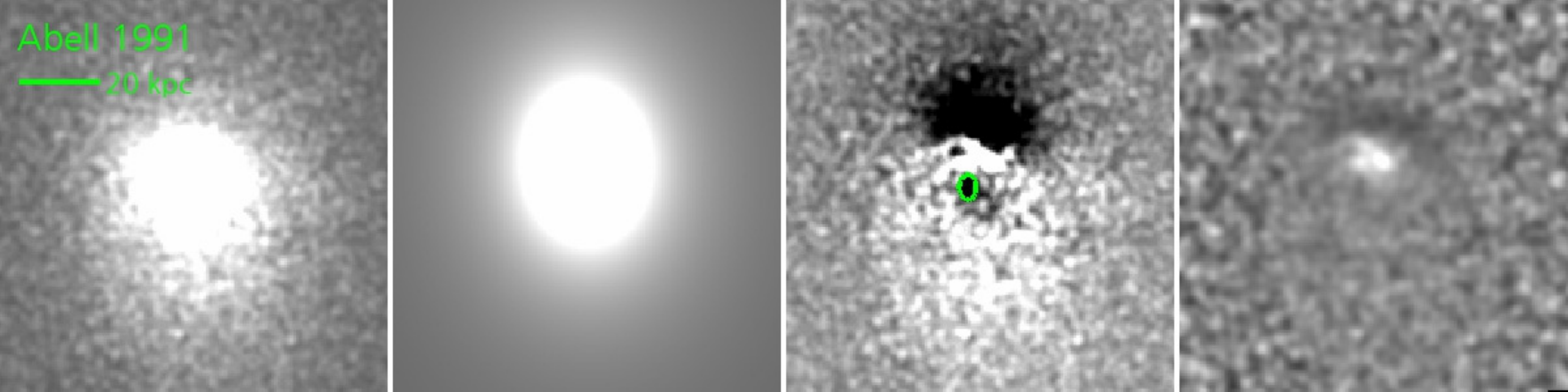}
\centering
\caption{
     Example of cold front targets, Abell 1991. Smoothed raw image (left), 
     $\beta$-model (middle left), residual image after subtracting
     $\beta$-model (middle right), and unsharp masked image (right).
\label{fig:allspec12}}
\end{figure*} 

In our analysis, we find that asymmetric gas distributions (i.e.,
cold fronts) can produce false detections of X-ray cavities if we use
a symmetric $\beta$-model.  In Figure 5, we present such a case, Abell
1991, for which \cite{Dong2010} detected the presence of two cavities
(see Figure 4. in \citealt{Dong2010}).  The distribution of the
diffuse emission is somewhat vertically asymmetric in the raw image
with a sharp drop in the north and a shallow decrease in the south.
As seen in the residual image after subtracting the $\beta$-model, we
detected only one cavity in the south (middle right panel). There seems
to be one more depression in the north but it may be still caused by
the asymmetric X-ray gas distribution since the symmetric
$\beta$-model can produce a false depression by over-subtracting the 
diffuse emission, respectively, in the north and
south in the image. Thus, we conclude that the depression in the
north may not be a real cavity, and that the fitting result may not be
reliable for such targets with a strong asymmetric gas distribution.
Nevertheless, as shown in the south, some targets show real X-ray
cavities in the opposite part of sharp drop, where the diffuse
emission is under-subtracted. Thus, the depression in the south is not
likely a false detection.

Point sources (AGNs) in the center of diffuse emission may also lead
to false detections.  For example, in the presence of a strong point
source at the center, the $\beta$-model tends to over-subtract diffuse
emission around the point source, producing a false depression (see
e.g., Abell 2029 and UGC12491 in Figure 4). For these cases, we did
not classify them as real cavities.  As presented in Figure 4, most
of targets with no cavity detection present these two features, i.e.,
asymmetric distribution or over-subtraction around the point
source. In these objects, it is not clear whether true cavities are
present.

\subsection{Gas temperature determination}\label{section:analysis}

To investigate the effect of the environments on X-ray cavities, we
quantified the environments of each target in the sample. Since
environments can be represented by various physical parameters, e.g.,
velocity dispersion of galaxies, X-ray gas temperature, and virial
mass, we reviewed literatures for measured velocity dispersions and
gas temperatures of each target.  However, the reported values are
often measured from different methods and incomplete for our sample.

To perform a uniform analysis and minimize systematic uncertainties,
we directly measured the gas temperatures within $\rm 0.15\ R_{2500}$
based on the X-ray data.  Since our targets cover a large redshift
range (from 0.001 to 2.7), it is difficult to measure $\rm R_{2500}$
for all targets since, for example, the $\rm R_{2500}$ of low-redshift
targets is far larger than the field of view (FOV) of Chandra images.  After testing
with various radii, we determined that $\rm 0.15\ R_{2500}$ is the
optimal radius for the majority of targets in the sample. For several
lower redshift targets, for which even $\rm 0.15\ R_{2500}$ is larger
than the FOV, we adopted a smaller radius: $\rm 0.10\ R_{2500}$ for
M84 and $\rm 0.05\ R_{2500}$ for NGC 4472, M87, NGC 4552, NGC 4636,
and NGC 4649, respectively.

The $\rm 0.15\ R_{2500}$ was calculated from the temperature-radius
relation (Equation 12 of \citealt{Vikhlinin2006}), using an initial
temperature value. Since the $\rm R_{2500}$ varies depending on gas
temperatures, we estimated $\rm 0.15\ R_{2500}$ by iterating radius
determination until the estimated temperature becomes consistent
within 10\%.  Note that the uncertainty of gas temperature is large
since we calculate $\rm 0.15\ R_{2500}$ from gas temperature derived
within $\rm 0.15\ R_{2500}$ while we used the temperature-radius
relation derived for $\rm R_{2500}$.

The gas temperature derived from $\rm 0.15\ R_{2500}$  differs on average by 20\% compared to that derived from $\rm R_{2500}$ \citep{Vikhlinin2006}.
The method for gas temperature measurement is same as described in \S2, except for the extraction radius. 
We estimated the uncertainties by combining 1 $\sigma$ fitting error
and 20\% systematic error to account for the temperature variation due
to the adopted size of the aperture (i.e., $\rm 0.15\ R_{2500}$
vs. $\rm R_{2500}$).  If a smaller radius (i.e., $< 0.15\ R_{2500}$)
was used for gas temperature measurements, we adopted 30\% higher
uncertainty, which is a maximum temperatures difference between $\rm
0.05\ R_{2500}$ $\rm 0.15\ R_{2500}$ \citep[see][]{Vikhlinin2006}.  If
there is no reliable fitting error when $\chi^2$ is larger than 2, we
adopted the maximum fitting error from other targets. Gas temperatures
and \HI\ column densities are presented in Table 1. 
 Note that for individual sources, the uncertainty of the estimated temperature 
can be larger than 20-30\% since $R_{2500}$ can be underestimated due to the degeneracy between
temperature and radius.

\begin{longtable}{lcccc}
\caption{Cavity properties}\\
\hline

\multicolumn{1}{c}{Object} &  \multicolumn{1}{c}{$r_{a}$} & \multicolumn{1}{c}{$r_{b}$} &
 \multicolumn{1}{c}{Area} & \multicolumn{1}{c}{Distance} \\

 &  \multicolumn{1}{c}{(kpc)} & \multicolumn{1}{c}{(kpc)} & \multicolumn{1}{c}{(kpc$^{2}$)} & 
 \multicolumn{1}{c}{(kpc)}\\

\multicolumn{1}{c}{(1)} & \multicolumn{1}{c}{(2)}  & \multicolumn{1}{c}{(3)} &
\multicolumn{1}{c}{(4)} & \multicolumn{1}{c}{(5)} \\
\hline

\endfirsthead

\multicolumn{5}{c}
{\tablename\ \thetable\ -- Continued \textit{}} \\
\\
\hline

\multicolumn{1}{c}{Object} &  \multicolumn{1}{c}{$r_{a}$} & \multicolumn{1}{c}{$r_{b}$} &
 \multicolumn{1}{c}{Area} & \multicolumn{1}{c}{Distance} \\
 &  \multicolumn{1}{c}{(kpc)} & \multicolumn{1}{c}{(kpc)} & \multicolumn{1}{c}{(kpc$^{2}$)} & 
 \multicolumn{1}{c}{(kpc)}\\

\multicolumn{1}{c}{(1)} & \multicolumn{1}{c}{(2)}  & \multicolumn{1}{c}{(3)} &
\multicolumn{1}{c}{(4)} & \multicolumn{1}{c}{(5)} \\\hline

\endhead
\endfoot
\hline
\multicolumn{5}{l}{\scriptsize{\hspace{2ex} Note -- Col. (1): Object name. Col. (2): Semimajor axis. Col. (3): }} \\
\multicolumn{5}{l}{\scriptsize{Semiminor axis. Col. (4): Area of X-ray cavity. Col. (5): Distance  }} \\
\multicolumn{5}{l}{\scriptsize{from X-ray gas emission center to X-ray cavity center. The border }} \\
\multicolumn{5}{l}{\scriptsize{divides the detected X-ray cavities from raw images (upper) and beta}} \\
\multicolumn{5}{l}{\scriptsize{-model subtracted images (lower).}} \\

\endlastfoot
Abell 426		&$	16.4	$&$	6.0	$&$	307.5	$&$	29.6	$	\\
		&$	6.9	$&$	4.6	$&$	99.2	$&$	6.5	$	\\
		&$	6.9	$&$	5.6	$&$	120.5	$&$	10.2	$	\\
		&$	6.3	$&$	5.5	$&$	108.6	$&$	21.0	$	\\
		&$	9.1	$&$	7.3	$&$	208.1	$&$	35.7	$	\\
											
NGC 1316		&$	0.6	$&$	0.4	$&$	0.6	$&$	0.8	$	\\
		&$	0.8	$&$	0.8	$&$	2.0	$&$	1.2	$	\\
		&$	0.2	$&$	0.2	$&$	0.1	$&$	0.7	$	\\

2A 0335+096		&$	9.6	$&$	5.8	$&$	174.8	$&$	30.0	$	\\
		&$	6.3	$&$	4.3	$&$	83.9	$&$	28.4	$	\\
		&$	9.6	$&$	5.6	$&$	168.0	$&$	17.7	$	\\
											
Abell 478		&$	4.3	$&$	3.9	$&$	53.4	$&$	9.6	$	\\
		&$	4.2	$&$	4.1	$&$	54.6	$&$	11.3	$	\\
											
MS 0735.6+7421		&$	82.5	$&$	72.2	$&$	18714.7	$&$	167.4	$	\\
		&$	92.1	$&$	79.2	$&$	22921.3	$&$	139.5	$	\\
											
Hydra A		&$	16.4	$&$	11.2	$&$	574.6	$&$	23.8	$	\\
		&$	15.3	$&$	9.6	$&$	461.0	$&$	24.8	$	\\
											
RBS 797		&$	13.5	$&$	9.8	$&$	414.3	$&$	25.0	$	\\
		&$	16.2	$&$	13.3	$&$	677.8	$&$	30.1	$	\\
											
M84		&$	1.5	$&$	1.3	$&$	6.0	$&$	1.8	$	\\
		&$	1.7	$&$	1.6	$&$	8.3	$&$	2.1	$	\\
											
M87		&$	0.7	$&$	0.6	$&$	1.3	$&$	2.2	$	\\
		&$	1.0	$&$	0.6	$&$	2.1	$&$	2.4	$	\\
		&$	1.0	$&$	0.5	$&$	1.6	$&$	1.2	$	\\
		&$	0.6	$&$	0.4	$&$	0.7	$&$	0.7	$	\\
		&$	1.0	$&$	0.5	$&$	1.5	$&$	1.8	$	\\
		&$	1.1	$&$	0.5	$&$	1.9	$&$	1.6	$	\\
		&$	1.0	$&$	0.8	$&$	2.5	$&$	3.3	$	\\
											
NGC 4552		&$	0.2	$&$	0.1	$&$	0.1	$&$	0.2	$	\\
		&$	0.2	$&$	0.1	$&$	0.1	$&$	0.2	$	\\
											
NGC 4636		&$	2.2	$&$	1.4	$&$	9.5	$&$	3.8	$	\\
		&$	2.2	$&$	1.4	$&$	9.5	$&$	3.8	$	\\
		&$	2.4	$&$	1.5	$&$	11.3	$&$	3.7	$	\\
		&$	2.3	$&$	1.2	$&$	8.9	$&$	3.5	$	\\
											
Centaurus cluster		&$	2.9	$&$	1.5	$&$	13.5	$&$	3.5	$	\\
		&$	2.0	$&$	1.1	$&$	7.1	$&$	3.6	$	\\
		&$	1.2	$&$	0.9	$&$	3.3	$&$	3.6	$	\\
		&$	2.7	$&$	1.3	$&$	11.3	$&$	5.7	$	\\
											
HCG 62		&$	5.4	$&$	4.6	$&$	77.7	$&$	8.6	$	\\
		&$	5.7	$&$	5.6	$&$	98.9	$&$	11.7	$	\\
											
NGC 5044		&$	1.7	$&$	1.0	$&$	5.1	$&$	4.5	$	\\
		&$	4.1	$&$	3.4	$&$	43.6	$&$	6.5	$	\\
		&$	2.9	$&$	2.5	$&$	22.8	$&$	8.1	$	\\
											
Abell 3581		&$	3.0	$&$	2.4	$&$	23.2	$&$	5.4	$	\\
		&$	4.2	$&$	2.7	$&$	36.3	$&$	6.7	$	\\
											
NGC 5813		&$	1.0	$&$	0.7	$&$	2.2	$&$	1.6	$	\\
		&$	1.0	$&$	0.8	$&$	2.5	$&$	1.5	$	\\
		&$	4.8	$&$	4.5	$&$	68.4	$&$	7.6	$	\\
		&$	4.6	$&$	4.1	$&$	59.3	$&$	7.6	$	\\
											
Abell 2052		&$	4.3	$&$	3.0	$&$	40.5	$&$	13.9	$	\\
		&$	6.7	$&$	3.7	$&$	77.9	$&$	6.0	$	\\
		&$	5.7	$&$	3.0	$&$	54.4	$&$	7.5	$	\\
		&$	8.6	$&$	4.6	$&$	124.2	$&$	8.5	$	\\
		&$	7.1	$&$	4.2	$&$	93.5	$&$	11.0	$	\\
		&$	12.2	$&$	5.7	$&$	218.1	$&$	18.1	$	\\
											
3C 320		&$	16.3	$&$	12.9	$&$	660.5	$&$	28.5	$	\\
		&$	15.1	$&$	12.5	$&$	592.2	$&$	26.8	$	\\
											
Cygnus A		&$	21.1	$&$	17.5	$&$	1163.1	$&$	37.2	$	\\
		&$	11.7	$&$	7.6	$&$	279.0	$&$	24.3	$	\\
		&$	6.5	$&$	3.4	$&$	69.4	$&$	16.0	$	\\
		&$	28.1	$&$	19.2	$&$	1698.0	$&$	40.2	$	\\
											
PKS 2153-69		&$	11.6	$&$	7.1	$&$	259.7	$&$	14.8	$	\\
		&$	10.8	$&$	7.5	$&$	253.4	$&$	15.9	$	\\
											
3C 444		&$	61.4	$&$	46.6	$&$	9000.7	$&$	83.5	$	\\
		&$	50.8	$&$	43.0	$&$	6873.6	$&$	75.3	$	\\
\hline											
Abell 85		&$	8.1	$&$	5.4	$&$	136.8	$&$	17.1	$	\\
											
ZwCl 0040+2404		&$	14.5	$&$	11.7	$&$	533.3	$&$	14.0	$	\\
											
ZwCl 0104+0048		&$	6.6	$&$	6.6	$&$	135.7	$&$	--	$	\\
											
NGC 533		&$	1.4	$&$	0.9	$&$	4.0	$&$	1.6	$	\\
		&$	1.3	$&$	1.1	$&$	4.4	$&$	1.8	$	\\
											
Abell 262		&$	3.7	$&$	2.5	$&$	29.4	$&$	3.7	$	\\
		&$	4.2	$&$	3.2	$&$	42.2	$&$	6.4	$	\\
											
MACS J0242.5-2132		&$	12.7	$&$	7.0	$&$	279.7	$&$	12.1	$	\\
MACS J0242.5-2132		&$	23.6	$&$	17.0	$&$	1261.2	$&$	28.7	$	\\
											
Abell 383		&$	16.0	$&$	7.4	$&$	371.0	$&$	8.2	$	\\
											
AWM 7		&$	2.7	$&$	2.1	$&$	17.7	$&$	3.9	$	\\
		&$	3.1	$&$	2.3	$&$	22.3	$&$	3.5	$	\\
											
MACS J0329.6-0211		&$	17.4	$&$	9.4	$&$	513.0	$&$	18.2	$	\\
		&$	24.2	$&$	16.4	$&$	1247.9	$&$	21.3	$	\\
											
NGC 1399		&$	0.2	$&$	0.2	$&$	0.1	$&$	0.2	$	\\
		&$	0.2	$&$	0.1	$&$	0.1	$&$	0.3	$	\\
											
NGC 1404		&$	0.5	$&$	0.4	$&$	0.7	$&$	0.7	$	\\
		&$	0.4	$&$	0.3	$&$	0.4	$&$	0.4	$	\\
											
RXC J0352.9+1941		&$	8.0	$&$	5.9	$&$	148.2	$&$	9.3	$	\\
		&$	7.9	$&$	4.3	$&$	107.4	$&$	10.8	$	\\
											
MACS J0417.5-1154		&$	13.8	$&$	9.4	$&$	406.8	$&$	10.9	$	\\
		&$	38.0	$&$	29.7	$&$	3543.8	$&$	76.4	$	\\
											
RX J0419.6+0225		&$	1.2	$&$	0.9	$&$	3.6	$&$	1.7	$	\\
											
EXO 0423.4-0840		&$	9.9	$&$	5.9	$&$	184.4	$&$	14.0	$	\\
		&$	7.1	$&$	4.8	$&$	106.5	$&$	10.4	$	\\
											
Abell 496		&$	6.5	$&$	4.7	$&$	97.1	$&$	8.8	$	\\
		&$	5.9	$&$	5.6	$&$	103.2	$&$	6.3	$	\\
											
RXC J0439.0+0520		&$	35.6	$&$	25.0	$&$	2796.2	$&$	25.3	$	\\
											
MS 0440.5+0204		&$	7.2	$&$	5.9	$&$	133.1	$&$	15.7	$	\\
		&$	21.8	$&$	9.5	$&$	650.7	$&$	17.7	$	\\
											
 MACS J0744.9+3927 		&$	45.3	$&$	28.6	$&$	4062.3	$&$	43.6	$	\\
		&$	37.1	$&$	21.0	$&$	2439.5	$&$	28.6	$	\\
											
PKS 0745-19 		&$	9.9	$&$	8.8	$&$	273.5	$&$	18.4	$	\\
		&$	11.6	$&$	8.5	$&$	310.7	$&$	9.3	$	\\
											
ZwCl 0949+5207		&$	16.1	$&$	13.9	$&$	699.6	$&$	21.9	$	\\
		&$	18.5	$&$	12.0	$&$	696.1	$&$	18.1	$	\\
											
ZwCl 1021+0426		&$	13.5	$&$	7.1	$&$	300.0	$&$	33.0	$	\\
											
RXC J1023.8-2715		&$	52.9	$&$	27.8	$&$	4623.6	$&$	38.5	$	\\
		&$	36.6	$&$	21.2	$&$	2441.9	$&$	61.4	$	\\
											
Abell 1068		&$	10.9	$&$	8.7	$&$	299.3	$&$	22.7	$	\\
											
Abell 1204		&$	22.1	$&$	17.4	$&$	1212.4	$&$	23.8	$	\\
											
NGC 4104		&$	0.9	$&$	0.6	$&$	1.7	$&$	--	$	\\
											
NGC 4472		&$	0.3	$&$	0.2	$&$	0.2	$&$	0.5	$	\\
		&$	0.3	$&$	0.2	$&$	0.2	$&$	0.4	$	\\
											
Abell 1689		&$	79.9	$&$	71.1	$&$	17843.4	$&$	109.6	$	\\
		&$	63.9	$&$	46.4	$&$	9314.9	$&$	67.7	$	\\
											
RX J1350.3+0940		&$	8.1	$&$	7.8	$&$	199.3	$&$	9.1	$	\\
											
ZwCl 1358+6245		&$	31.4	$&$	24.9	$&$	2451.7	$&$	33.4	$	\\
		&$	22.3	$&$	20.3	$&$	1420.4	$&$	45.5	$	\\
											
Abell 1835		&$	16.9	$&$	9.9	$&$	527.8	$&$	14.3	$	\\
		&$	5.8	$&$	5.6	$&$	102.3	$&$	11.0	$	\\
											
MACS J1423.8+2404		&$	27.1	$&$	12.7	$&$	1082.3	$&$	36.1	$	\\
											
Abell 1991		&$	4.5	$&$	3.3	$&$	46.3	$&$	8.3	$	\\
											
RXC J1504.1-0248		&$	13.6	$&$	4.5	$&$	191.6	$&$	11.9	$	\\
		&$	36.6	$&$	15.6	$&$	1789.8	$&$	23.5	$	\\
											
NGC 5846		&$	1.1	$&$	0.5	$&$	1.7	$&$	0.8	$	\\
		&$	0.5	$&$	0.4	$&$	0.6	$&$	0.5	$	\\
											
MS 1512-cB58  		&$	43.4	$&$	18.4	$&$	2507.1	$&$	73.0	$	\\
											
 RXC J1524.2-3154		&$	19.5	$&$	8.9	$&$	545.1	$&$	13.9	$	\\
		&$	5.4	$&$	4.1	$&$	70.3	$&$	16.3	$	\\
		&$	11.1	$&$	6.4	$&$	223.1	$&$	23.3	$	\\
											
RX J1532.9+3021		&$	31.3	$&$	16.8	$&$	1653.4	$&$	32.7	$	\\
		&$	29.9	$&$	14.5	$&$	1359.1	$&$	22.7	$	\\
											
Abell 2199		&$	8.0	$&$	6.2	$&$	155.0	$&$	19.7	$	\\
		&$	8.6	$&$	7.0	$&$	187.8	$&$	22.7	$	\\
											
Abell 2204		&$	10.0	$&$	9.4	$&$	295.2	$&$	13.0	$	\\
											
NGC 6338		&$	3.8	$&$	2.3	$&$	27.1	$&$	3.2	$	\\
		&$	3.7	$&$	2.5	$&$	28.5	$&$	4.3	$	\\
											
MACS J1720.3+3536		&$	17.6	$&$	16.8	$&$	929.8	$&$	25.4	$	\\
		&$	20.7	$&$	16.8	$&$	1089.1	$&$	27.5	$	\\
											
MACS J1931.8-2634		&$	19.6	$&$	9.0	$&$	551.3	$&$	30.3	$	\\
											
MACS J2046.0-3430		&$	14.4	$&$	7.2	$&$	328.2	$&$	--	$	\\
											
MACS J2229.7-2755		&$	8.8	$&$	7.2	$&$	199.5	$&$	12.3	$	\\
		&$	11.8	$&$	9.0	$&$	333.7	$&$	18.5	$	\\
											
Sersic 159-03		&$	15.3	$&$	12.3	$&$	592.9	$&$	29.5	$	\\
		&$	6.2	$&$	4.4	$&$	86.4	$&$	13.5	$	\\
		&$	20.5	$&$	16.3	$&$	1052.7	$&$	21.5	$	\\
											
Abell 2597		&$	7.3	$&$	6.0	$&$	138.9	$&$	21.7	$	\\
		&$	3.1	$&$	2.9	$&$	28.7	$&$	9.4	$	\\
		&$	8.7	$&$	6.6	$&$	179.6	$&$	21.6	$	\\
		&$	5.1	$&$	3.0	$&$	48.2	$&$	20.6	$	\\
		&$	14.0	$&$	9.6	$&$	422.6	$&$	17.4	$	\\
											
Abell 2626		&$	19.9	$&$	12.4	$&$	775.5	$&$	18.1	$	
\end{longtable}

\section{Result} \label{section:result}

\subsection{Cavity detection} \label{section:cavity detection}

Using the raw images and applying the $\beta$-model, we detected 148 cavities from 69 targets. 
Out of the 133 targets in our sample with sufficient X-ray photons, $\sim$ 52\% shows cavities. 
Among them, 18, 38, and 13 targets show one, two and more than two cavities, respectively.  
In general, two cavities are detected as a pair on opposite sides of the center, except for a few cases.
Since X-ray cavities are believed to be formed by a radio jet, two bi-symmetric X-ray cavities are 
expected. However, the presence of more than two cavities in many targets may imply that
there were multiple radio events in the past. In the case of more than 2 cavities, 4 targets show 
one pair and one additional cavity (1, NGC 1316) while the other 9 targets show various geometry:
collinear two pairs (e.g., NGC 5813), two pairs with different direction (2, e.g., NGC 4636 \& Cygnus A), 
or a complicated structure (6, e.g.,  Abell 426 \& M87).

\begin{figure*}
\center
\includegraphics[width = 0.90\textwidth]{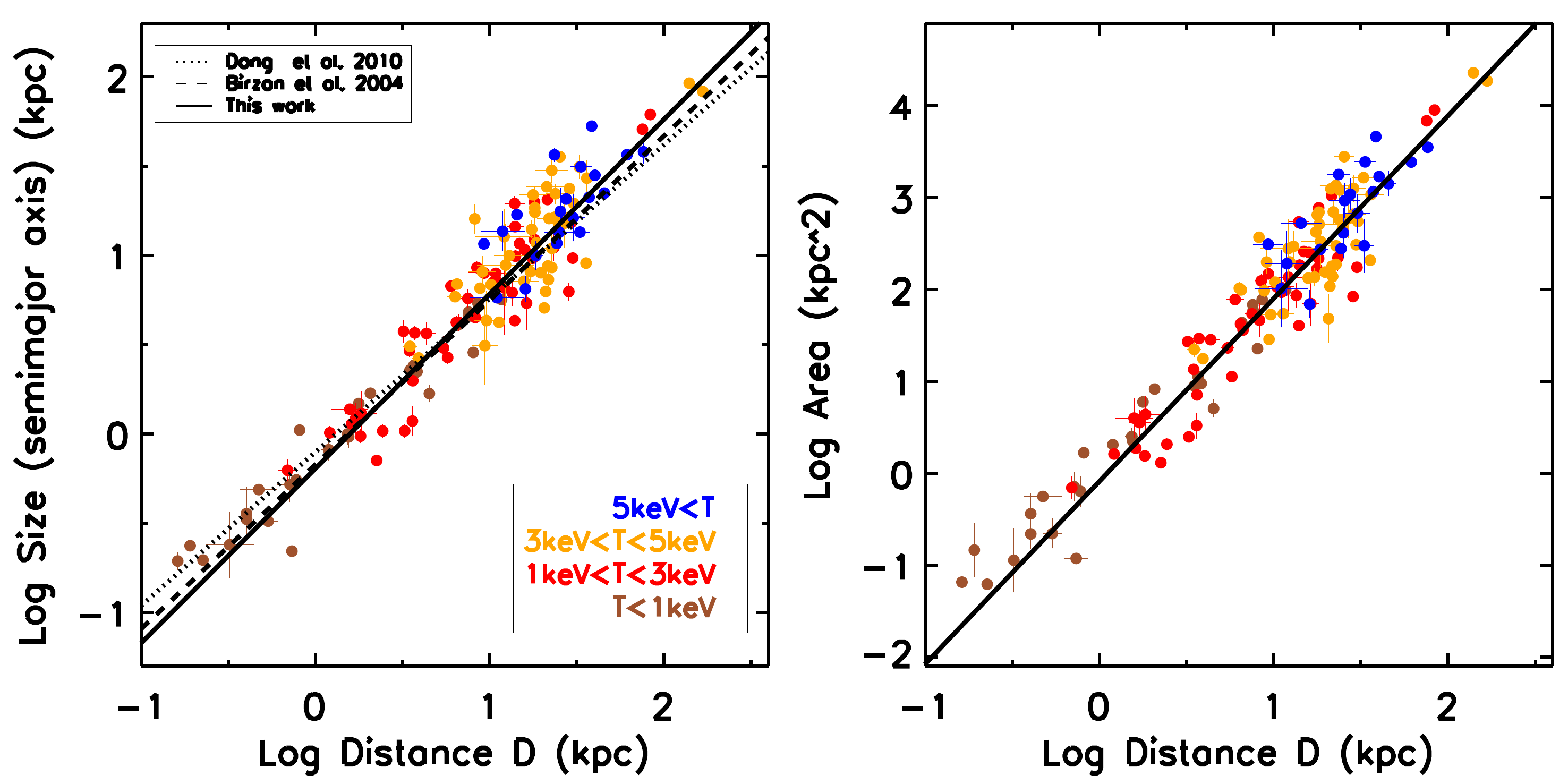}
\caption{
     Comparison between our result and previous results. Semimajor axis and distance comparison (left), and area and distance comparison (right). 
\label{fig:allspec1}}
\endcenter
\end{figure*}

\subsection{Cavity properties comparison}

 To investigate relations between cavity properties, we compared
  the cavity size in the semimajor axis (a), distance from the X-ray center (D), and area as
  shown in Figure 6, using the "FITEXY" method
  \citep{Tremaine2002,Park2012}.  To calculate uncertainties, we
  performed Monte Carlo simulations with 10,000 mock datasets by
  randomizing the error, and adopted the standard deviation as the
  uncertainty of the fitting results.
  
Over the large dynamic range covered by our sample, we confirm a
linear correlation of the cavity size with the distance from the X-ray
center,
\begin{equation}              
{\rm log}\ a= (0.97 \pm 0.02)\ {\rm log}\ D- 0.15\pm 0.02,
\end{equation}
as previous studies reported \citep[][]{Birzan2004,Dong2010}. 
There are several explanations to understand this correlation in terms of cavity evolution such as 
buoyancy or momentum from AGN jet. 
By testing various models, \cite{Diehl2008c} showed that
magnetically-dominated cavities can be well explained by current-dominated magnetohydrodynamic jet models
\citep{Li2006,Nakamura2006,Nakamura2007}).

Since the cavity shape is generally elliptical, cavity area may better
represent the cavity property than the size in the semimajor- or
semiminor- axes.  In addition, to estimate the total energy emitted
from a radio jet, the volume of cavities is required based on
additional assumptions (i.e., cavity shape).  Instead, cavity area can
be measured from the projected image and compared to the distance.  As
shown in right panel of Figure 6, we find a similar correlation between the cavity
area and the distance from the center
\begin{equation}              
{\rm log}\ Area= (1.94 \pm 0.04)\ {\rm log}\ D+ 0.02\pm 0.05.
\end{equation}
The scatter of the area-distance relation is similar to that of the size-distance relation (0.155$\pm$0.010 dex and 
0.163$\pm$0.011 dex, respectively, in distance unit).
For the rest of paper, we will use the area as the representative of the cavity size.
In Figure 6, gas temperatures are denoted with different colors.
Of interest, it seems that high temperature systems tend to have larger cavities. 
However, the cavity size correlates with the distance regardless of gas temperature, suggesting 
that the formation and evolution of the X-ray cavities has little dependence on environment.

\subsection{X-ray cavity and gas temperature}\label{section: cavity temperature}

To investigate the environmental effects on X-ray cavities, we first
checked the fraction of X-ray cavity detection as a function of X-ray
gas temperatures in Figure 7.  At each temperature bin, the number of
total targets and the number of targets with cavity detection are
represented, respectively, with the unfilled- and
filled-histogram. The detection rate of X-ray cavities is on
  average 49\% while the rate varies slightly in individual
  temperature bins. We do not find a significant change of the
  detection rate except for the very high temperature bins that
  contain small number of targets, suggesting that no strong
  environmental effects on the presence of X-ray cavities.

In contrast, there appears to be a trend between cavity size and
environments.  By comparing cavity sizes to gas temperatures (Figure
8), we find that the cavity size increases with gas temperatures.
Since many targets at given gas temperature have multiple cavities
with various sizes, the relation between the cavity size and gas
temperature is not strong.  Nonetheless, there is a clear trend
between them, suggesting that more massive systems possess larger size
cavities.  However, to confirm the correlation, we should look at the
selection effect carefully, which will be discussed in detail in \S
5.1.\\ 

\begin{figure}
\includegraphics[width = 0.4\textwidth]{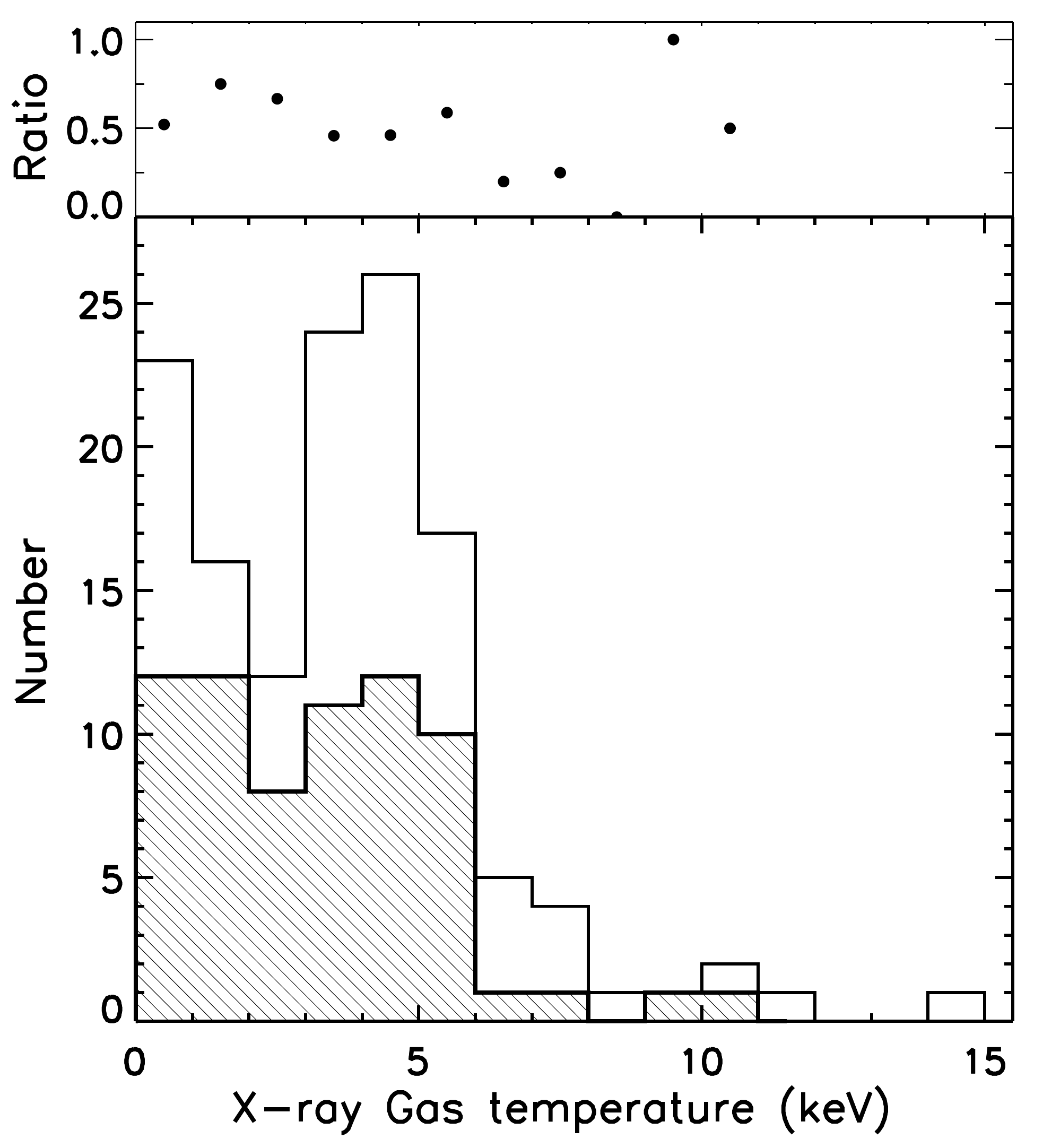}
\caption{
  X-ray cavity detection on temperature distribution. Total number (unfilled) and X-ray cavity detection
  number (filled) are represented.
  \label{fig:allspec1}}
\end{figure}

\begin{figure}
\includegraphics[width = 0.40\textwidth]{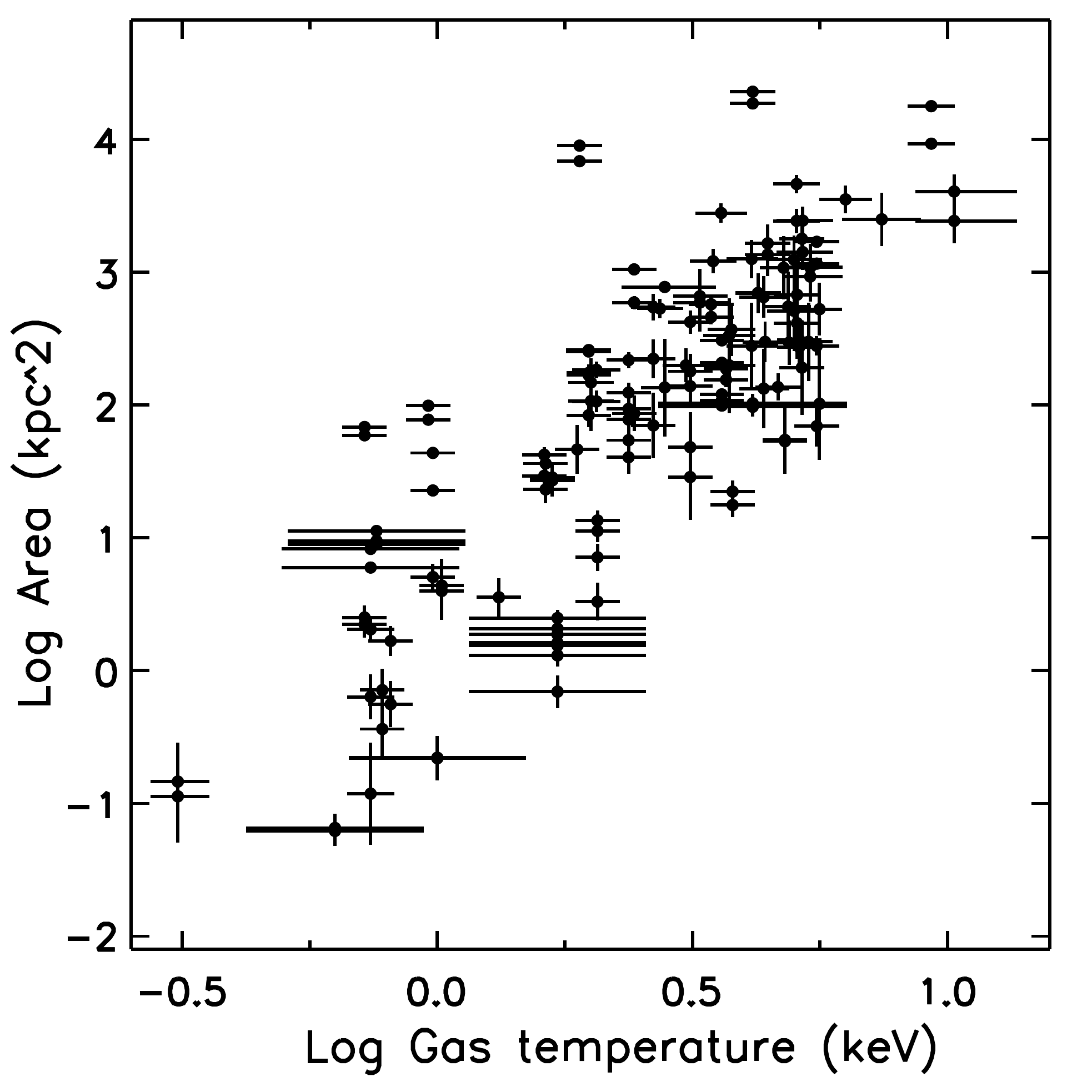}
\caption{
  Cavity size as a function of the gas temperature of each targets. 
\label{fig:allspec11}}
\end{figure}

\section{Discussion} \label{section:discussion}
\subsection{Dependence on environments}\label{section:discussion}

Previous studies reported that the X-ray cavity detection rate is
$\sim$2/3 for cool-core clusters \citep[14 out of 20;][]{Dunn2006},
$\sim$1/2 for galaxy groups \citep[26 out of 51;][]{Dong2010}, and
$\sim$1/4 for elliptical galaxies \citep[24 out of
  104;][]{Nulsen2009}.  At face value these results seem to suggest
that the detection rate of X-ray cavity increases with galaxy density,
indicating environment effects on the formation of X-ray cavities.

As pointed in \S 2, however, consistent criteria for sample selection
and uniform analysis is required to constrain the detection rate and
environment effect.  By investigating the targets in the previous
studies, we find that many targets, for which X-ray cavities were
detected, do not have sufficient X-ray photons.  For example, 17
targets of \cite{Dong2010} satisfy our sample selection criteria while
the other 38 targets have insufficient photon counts.  Among
  those 17 targets, we found cavities in 11 targets while
  \cite{Dong2010} detected cavities in 14 targets.    
  Comparing with the results of
  \cite{Dunn2006}, we used all 20 objects in their sample and detected
  cavities in 16 objects. While \cite{Dunn2006} also detected cavities in 16
  targets, they did not find cavities in the same 16 targets we did. 
We found X-ray cavities in Abell 496, Abell 2204, PKS 0745-191, and AWM7, however, \cite{Dunn2006} did not.
  It may be partly due to the fact that \cite{Dunn2006} used
  single exposures per target, without combining all available
  exposures and that they did not perform $\beta$-modeling and unsharp masking analysis.
  On the other hand, \cite{Dunn2006} reported X-ray
  cavities in Abell 1795, Abell 2029, Abell 4059, and MKW 3s, while we classified these target as
  non-detections since they show over-subtraction in our
  analysis.  These examples demonstrate the difficulties of direct
  comparison among various studies in the literature.  We provide more
  detailed comparison of our work to others in \S 5.4.

In contrast, our results suggest that the cavity detection rate does
not show significant dependence on gas temperatures (see Figure 7),
indicating that the formation and evolution of the X-ray cavities may
be driven by the central engine (i.e., AGN) and affected by the
physical conditions at the center (i.e., gas density) rather than the
global properties (i.e., environments).

  We showed that larger X-ray cavities tend to be detected in the
  higher gas temperature systems.  Since the cavity grows in size and
  moves away from the center, as indicated by the size-distance
  relationship, we expect to see small cavities close to the center
  even in high gas temperature systems.  The lack of small cavities in
  high gas temperature systems may reflect selection effects.  To
  investigate the effect, we compared the cavity size and gas
  temperature as a function of redshift (Figure 9), finding that
  systems with high gas temperature tend to be more distant since 
  more massive and more luminous systems  with high gas temperature are rarer. 
  Also, we tend to detect
  larger cavities at higher redshift as shown in Figure 9.  This trend
  can be due to the limited spatial resolution of the Chandra data.
  Since it is difficult to identify a cavity with a size smaller than
  $\sim$1\arcsec with the $Chandra$ PSF, we could not detect small
  size cavities at high redshift.  To illustrate this, we denoted the
  detection limit for a cavity with a 1\arcsec\ radius size as a
  function of redshift with a solid line in Figure 9. The distribution
  of the detected cavities over the redshift range is consistent with
  the detection limit, indicating that the lack of small cavities at
  high redshift is due to the fixed spatial resolution in Chandra
  images.

On the other hand, we tend to detect no large cavities in the systems
with lower gas temperature since the size of the diffuse X-ray
emitting region is intrinsically small.  
Since the cavities are detected as a form of depression as embedded in the diffuse X-ray gas,
the size of X-ray emitting region is the maximum limit of the cavity
size.  In Figure 10, we present the size of the X-ray halo with a grey
line, which represents $\rm R_{2500}$ as a function of gas
temperature while the thickness represents the range of $\rm R_{2500}$
at the redshift range 0.001<z<0.7, which is relevant for all targets in the sample
except for one target, MS 1512-cB58 at z = 2.723. This figure 
demonstrates the growth limit of
X-ray cavities. For example, 3C 444  ($\rm{\log\ (D /1 kpc) \sim 1.8}$) and MS
0735.6+7421 ($\rm{\log\ (D /1 kpc) \sim 2.2}$) contain relatively large cavities,
which are close to the growth limit and detected at the edge of the
X-ray distribution.  There are other cavities detected at the edge of
the X-ray distribution (e.g., M84 and NGC 4636).  However, these
target have much shallower X-ray images, hence, the size of the
observed X-ray distribution is much smaller than $\rm R_{2500}$.

In Figure 11, we demonstrate the two forbidden areas in the
size-temperature plane with the growth limit and the spatial
resolution limit.  Intrinsically the cavity size can not be larger
than the X-ray emitting region of individual targets, confining the
detected cavities below the growth limit, while Chandra image
resolution does not allow us to observationally detect small cavities
below the spatial resolution limit.

\begin{figure}
\includegraphics[width = 0.4\textwidth]{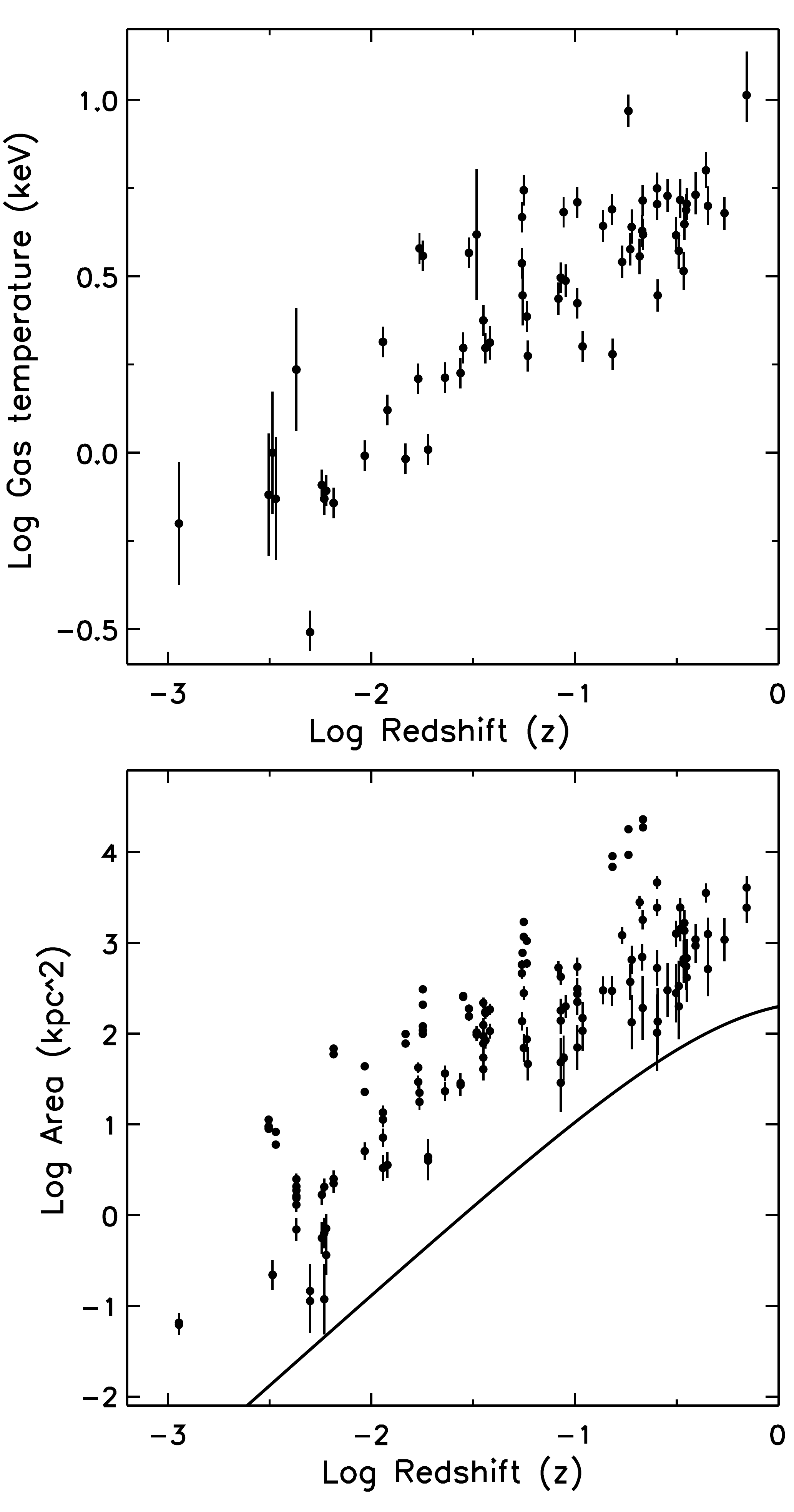}
\caption{
      Relation between gas temperature and redshift (top) and relation between 
      area of X-ray cavities and redshift (bottom). Solid line in the 
      bottom panel shows the area of a circle with a radius of 1\arcsec as a function of the redshift.
\label{fig:allspec1}}
\end{figure}

\begin{figure}
\center
\includegraphics[width = 0.4\textwidth]{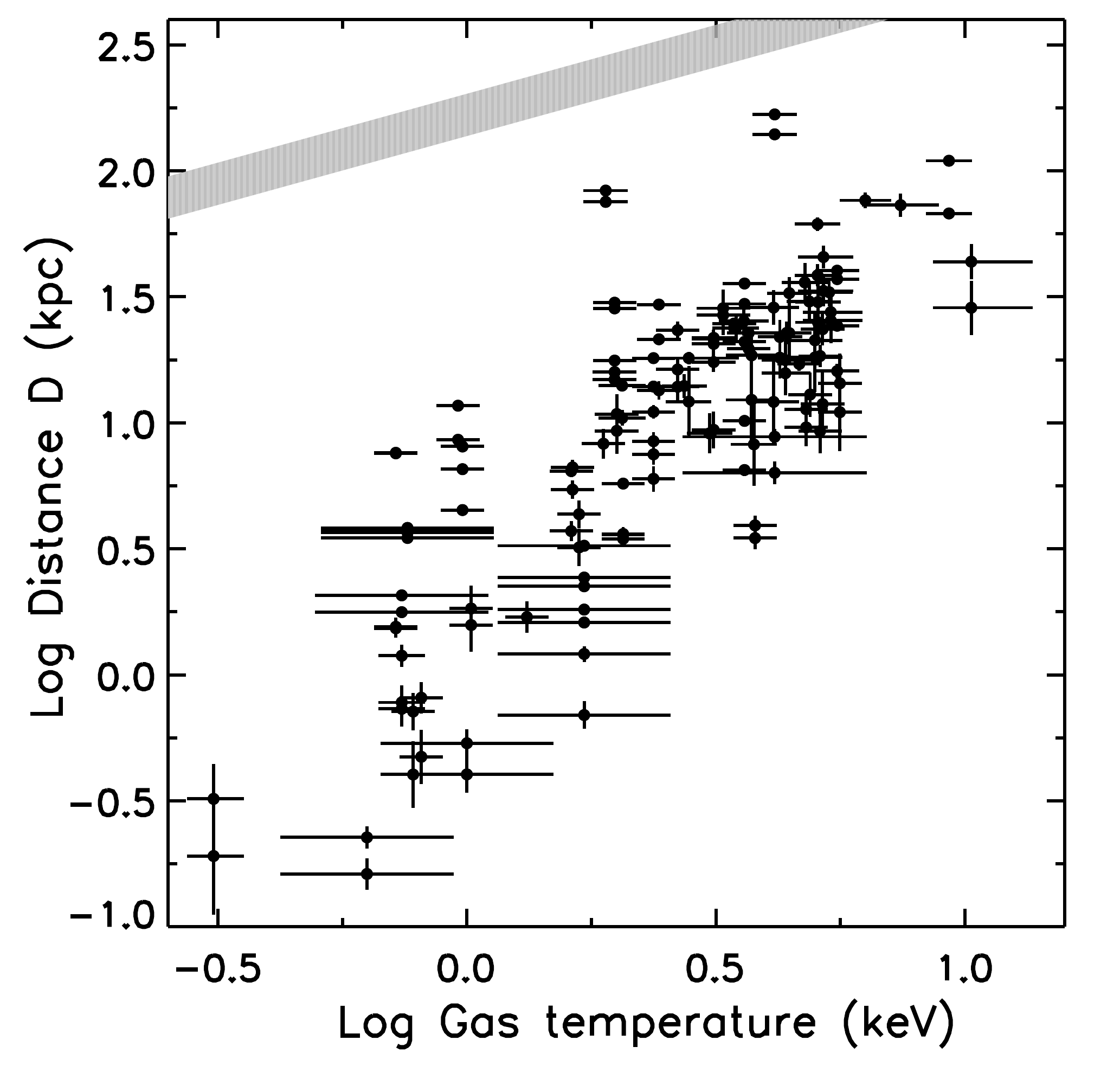}
\caption{
Distribution of gas temperature and cavity distance from the center. Thin gray region indicates
$\rm R_{2500}$ calculated from gas temperature for the redshift range (0.001<z<0.7).
\label{fig:allspec1}}
\endcenter
\end{figure}

\begin{figure}
\includegraphics[width = 0.45\textwidth]{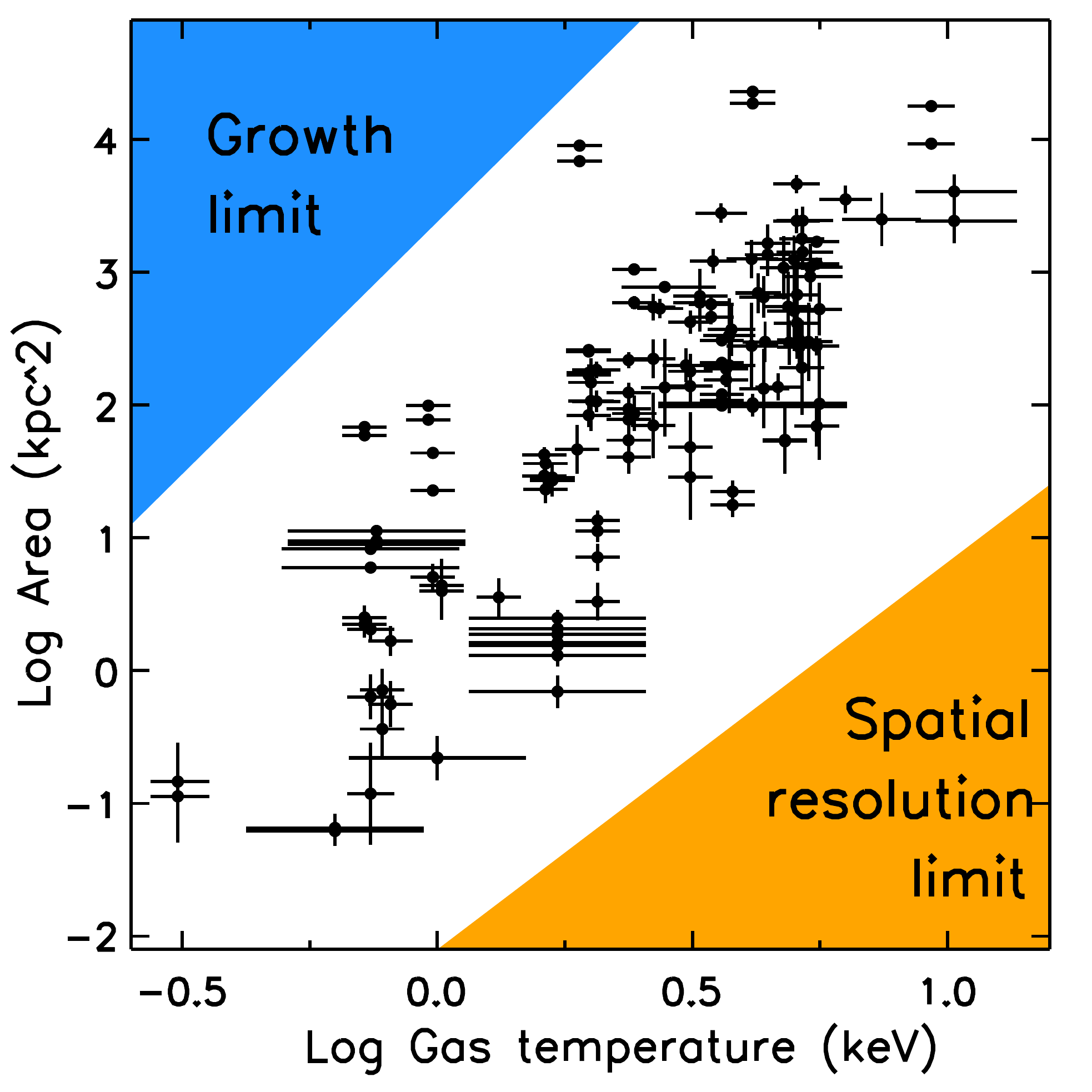}
\caption{
Same figure as Figure 8. Orange and blue areas mean
  spatial resolution limit area and growth limit area.
Black color shows the measurement from individual cavity. \label{fig:allspec1}}
\end{figure}

\subsection{Non-detection of X-ray cavity}\label{section:discussion}

While we detected X-ray cavities from 69 out of 133 targets, we
  did not find any clear evidence for X-ray cavities in the other 64
  targets (see \S 4).  Since the majority of non-detected targets have
  asymmetric distributions or over-subtraction issues, it is difficult
  to confirm the presence of cavities in these objects. This is the limitation of this work
  since we relied on symmetric $\beta$-modeling. The results from the $\beta$-modeling
  often present over-subtraction regions, which were classified as non-detections.  
For these targets, additional data such as radio can be an aid to reveal the true nature of diffuse X-ray emission. For example, \cite{Hodges2010, Hodges2010b} confirmed the presence of cavities by combining X-ray and radio data.

Another possible reason for non-detection is relatively weak X-ray depression. 
Previous work reported that cavity detectability strongly depends on the size and distance from the center, suggesting that large cavities with smaller distance can be detected easily \citep{Ensslin2002,Bruggen2009, Dong2010}. In other words, we tend to not detect small cavities far from the central region. In addition, \cite{Ensslin2002} showed that detectability also relies on the inclination angle of a cavity with respect to the plane of sky, showing that cavity depression is maximized when the inclination 
angle is zero. For these reasons, it is likely that we may have missed some cavities. These effects act as selection effects for our cavity detection and the observed size-distance relation presented in Figure 6.

We consider two additional scenarios for systems with no cavity
detections.  One scenario is that no X-ray cavity is present due to
the lack of AGN or jets.  If there are no radio jets or the radio jets
are too weak to push the hot X-ray gas, cavities may not form.  As a
possible AGN indicator, we investigate the presence of an X-ray point
source at the center of each object.  We found a central X-ray point
source for 33 out of 69 cavity-detected targets (48\%) and 33 out of
64 targets with no cavity detection (52\%), respectively, suggesting
that X-ray AGN fraction is similar between targets with/without
cavities. 
In the case of the cavity-detected targets without X-ray AGN, it is
possible that AGN is currently very weak or inactive while the black
hole was active in the past. In the case of no cavity targets with
X-ray AGNs, the AGN may not have strong jets to form cavities, or
cavities are present, but not detected due to the detection issues
(i.e., asymmetric distribution or over-subtraction). A detailed
investigation of the relationship between AGN and the presence of
cavities is beyond the scope of our current work.

Another possibility is that the radio jet was just launched or
launched a long time ago. If the radio jet is just launched, cavities
may be too small to resolve in the X-ray images.  If the radio jet
launched a long time ago, the X-ray cavities may have already moved
beyond the edge of the diffuse X-ray emission (see \S5.1)

To check this scenario, one can use multi-frequencies radio data.  In
detecting X-ray cavities directly, high-frequency data with high
resolution could reveal small X-ray cavities. However, due to the
energy loss through synchrotron emission, there could be no emission
at high-frequency if the jet was launched a long time ago.  In that
case, low-frequency observations might reveal extended features
\citep[e.g.,][]{Gitti2010,Giacintucci2011}. With the combination of
multi-frequency radio data, one can check for the presence of radio
jets at small and large scales. If there is no radio jet, it could
mean there was no AGN activity and thus no cavities as well.  On the
other hand, if there is the evidence of radio jets at small- or large-
scales, it could be good evidence to confirm the second scenario.
Furthermore, if the radio image shows multiple jets, 
it enables us to investigate the duty cycle of AGN activity \citep[e.g.,][]{Randall2011,Chon2012}.  
In means that we can study
in detail the period of AGN feedback duty cycle.  However, the current
lack of multi-frequency radio data for our targets makes such a study
not possible in the current paper.

We note that a scaling relation between jet power derived from the
cavities and radio luminosity has been found \citep{Birzan2004,
  Birzan2008, Cavagnolo2010, OSullivan2011a}. In other words, the jet
power can be estimated based on radio luminosity and used to predict
the presence of X-ray cavities. Thus, this may be another way to infer
the existence of cavities when they can not be directly detected in
X-ray images.
 
\subsection{The shape of X-ray cavities}\label{section:discussion}   

In Figure 12, we present a distribution of orientation angle of the
cavities with respect to the center (or jet direction).  We define the
orientation angle 0 degree if the cavity is radially elongated (along
with the jet direction; e.g., Cygnus A and Hydra A) while 90 degree
means that the cavity is elongated perpendicular to the jet direction
(e.g., NGC 4552 and HCG 62). Here, we only take the targets with a/b ratio
over 1.1 or under 0.9 to avoid a systematic uncertainty from circular shape cavity.
We find a bimodal distribution that cavities are elongated either parallel to the jet direction or
perpendicular to the jet direction, suggesting no significant trend
with respect to the jet direction. Note that since we did not fit
  the cavity with an ellipse model, and determined a representative model by
  visual inspection, the uncertainty of the direction of the
  elongation is relatively large. 

While various simulation results showed different orientations
\citep{Sijacki2006,Bruggen2007,Sijacki2007,Dubois2010,Gaspari2011,
  Li2014}, a recent study by \citet{Guo2015} investigated the
orientation of cavities with various input physical parameters,
suggesting that radial elongation increases with jet density,
velocity, and duration while it decreases with jet energy density and
radius, implying that the morphology of the cavities is closely
related to the jet properties.  \\

\begin{figure}
\center
\includegraphics[width = 0.40\textwidth]{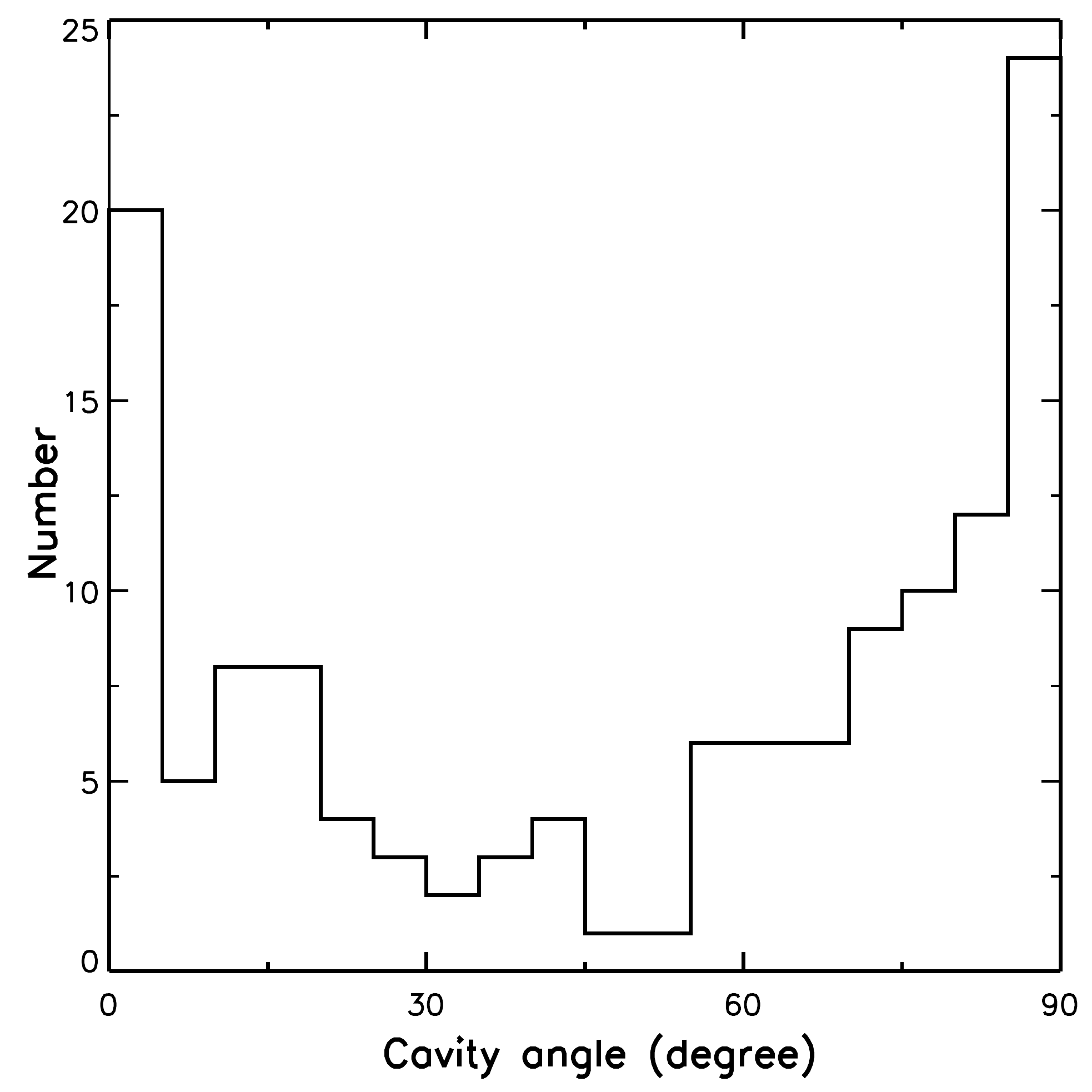}
\caption{
 Distribution of the orientation angle of X-ray cavities. 0 degree indicates radial elongation while 
90 degree means that major axis is perpendicular to the jet direction.}
\endcenter
\end{figure} 

\subsection{Comparison to previous works}\label{section:discussion}

  In this section, we compare our cavity detections with previous
  studies.  First, we used the initial sample of $\sim$800 targets,
  for which we obtained evidence for diffuse X-ray emission and
  performed a detailed X-ray count analysis (see \S 2), to find
  whether X-ray cavities were previously detected by other studies.
  By searching a number of papers cited for each target, we find that
  116 out of $\sim$800 targets were reported to possess X-ray cavities in
  the literature.

Among these 116 targets, 69 targets have Chandra images with
sufficient X-ray photons (i.e., larger than 4 counts per pixel at the
6-7 pixel annulus; see \S 2) while the X-ray image of the other 47
targets does not have sufficient photons, hence we did not perform the
detailed analysis for these targets.  The X-ray cavities of these 69
targets were identified by a number of works \citep[e.g.,][]{Birzan2004,
  Dunn2005, Dunn2006, Rafferty2006,Diehl2008c,
  Cavagnolo2010,Giacintucci2011, Birzan2012, Hlavacek2012,
  Russell2013, Panagoulia2014}.  When we compare them with our work,
we find that we independently detected cavities from 47 of these
targets. We did not detect X-ray cavities in 22 other targets that were
previously claimed to contain X-ray cavities. 
Since over-subtraction is shown in these targets, we
classified them as no-cavity detections (see Figure 4). For
these more complex objects, additional information such as radio flux
and morphology may help to confirm the existence of the cavities.
 Note that the presence of X-ray cavities have been confirmed using
radio data (e.g., IRAS 09104+4109, \citealt{OSullivan2012}; NGC 4261,
\citealt{OSullivan2011b}; Abell 2029, \citealt{Clarke2004}; and
Hercules A, \citealt{Nulsen2005b}).
However, to be consistent with other
over-subtraction targets, we left them as no-cavity detections based on X-ray data only. 

For the other 47 targets out of the 116 objects, X-ray cavities were
also reported in the literature \citep[e.g.,][]{Rafferty2006,
  Cavagnolo2010, Dong2010,Hodges2010,Hodges2010b,
  Lal2010,Machacek2010, OSullivan2010, Birzan2012, Panagoulia2014,
  Hlavacek2015}.  In our analysis, however, these targets were
excluded from the detailed $\beta$-modeling and further analysis since
the X-ray images did not satisfy the criterion of having sufficient
photons.  For example, \cite{Panagoulia2014} investigated a sample of
49 objects, finding X-ray cavities from 30 targets, while only 28
targets among their 49 objects were included in our analysis based on
the photon count criterion.
 As described in \S 2, we note that photon count is the most important parameter for detecting X-ray cavities in our analysis, thus deeper X-ray images are required for confirming cavities in these targets.

In our analysis of the total sample of 133 targets with sufficient
X-ray photons, we detected X-ray cavities for 69 objects.  While 47
objects among them were previously reported to have X-ray cavities, we
found new X-ray cavities for 22 additional objects for the first time.
In particular, 5 targets out of these 22 objects were studied by
previous works, however, no X-ray cavities were identified. They are
namely, AWM 7 \citep{Dunn2006, Birzan2012}, EXO 0423.4-0840, RXC J1504.1-0248 \citep{Birzan2012}, 
and Abell 2626 \citep{Panagoulia2014}. One of the reasons why we were able to detect
cavities in these objects is that we used deeper X-ray images by
combining individual exposures, which is the case for 4 targets. For
the other object, RXCJ 1504.1-0248, although the same X-ray data were used for both our and previous
works, we used the $\beta$-modeling while the previous works used raw
images and unsharp masking method (except for \citealt{Dong2010}). Thus,
the difference of the depth of the X-ray images and the analysis
method seems to explain the discrepancy.  

We find that the number of detected cavities for a given target is sometimes different from that of previous studies, 
presumably due to the use of different analysis methods. For example, we detected three cavities in NGC 5044 and
 NGC 1316, respectively, while \cite{David2009} detected four cavities in NGC 5044 and \cite{Lanz2010} found
 only one cavity in NGC 1316. \cite{Forman2007} and \cite{Wise2007} also detected cavities from Hydra A and M87, respectively,
 but the number of detected cavities in their studies is different from that of this work. In those studies, radio data were additionally used
 to confirm the presence of cavities, that were not classified as cavities with the X-ray data alone in our study,
 suggesting that our analysis and measurements of cavity properties are limited.
 A more statistical investigation of the various X-ray studies for cavities
 is beyond the scope of this paper. 

\begin{figure}
\includegraphics[width = 0.40\textwidth]{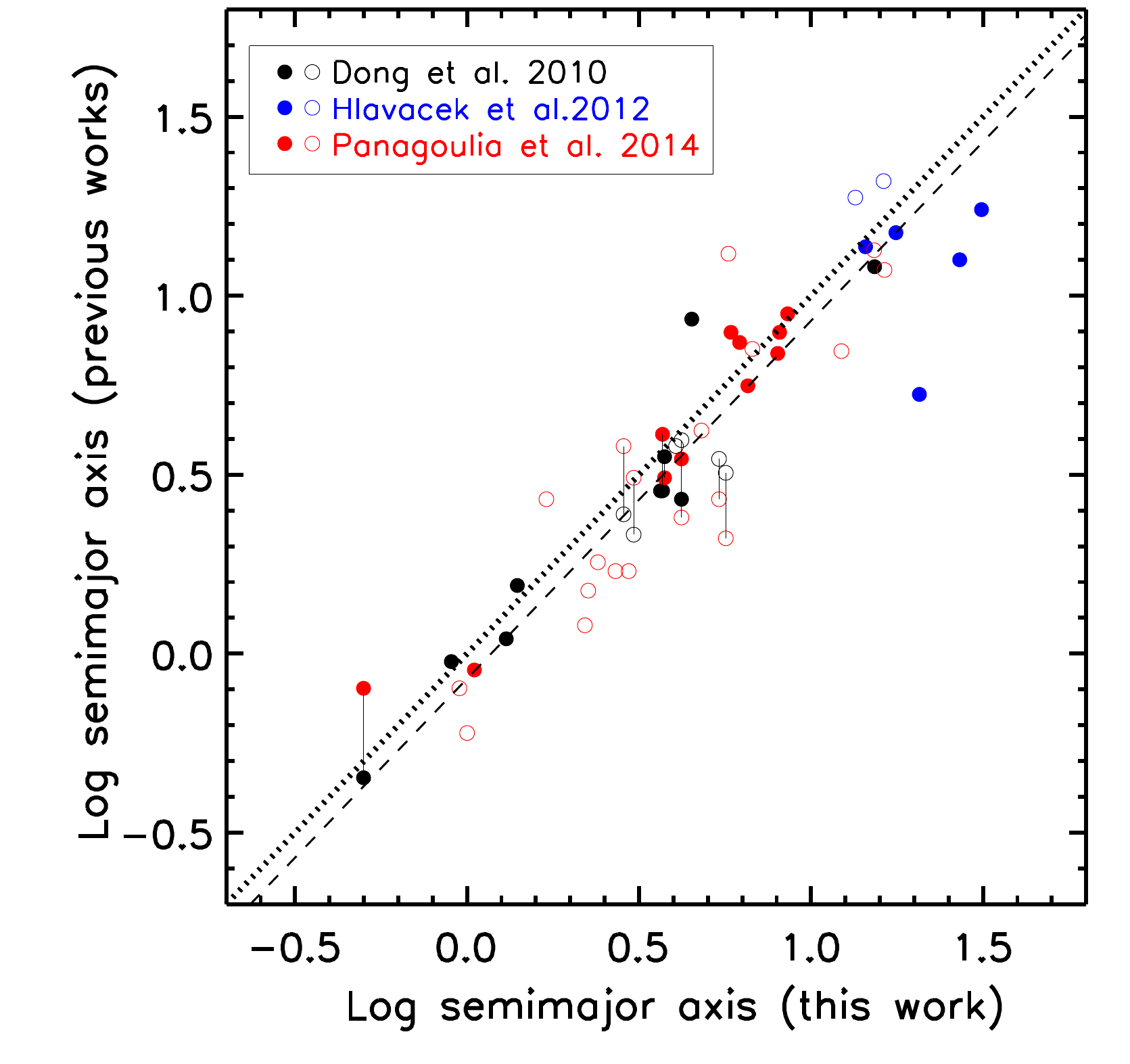}
\caption{
Comparison of cavity sizes. Black, blue and red points indicate the measurements by \citet{Dong2010,Hlavacek2012,Panagoulia2014}, respectively. 
Open and filled circle show cavity detected from raw image and based on the $\beta$-modeling.
Dotted and dashed lines denote one-to-one relationship and mean offset by 0.07 dex. The values show semi-major axis.}
\end{figure} 

We have compared the properties of the cavities detected in our study with
those in the literature for a consistency check. For this exercise, we used three studies
\citep{Dong2010,Hlavacek2012,Panagoulia2014} since they provide cavity
size as well as X-ray images, enabling us to crossmatch the cavities
detected in our study. We found 44 cavities in 24 targets that are common to these
other studies.
Among these targets, 9 cavities are detected by both
\cite{Dong2010} and \cite{Panagoulia2014}. By comparing the size of
these 44 cavities in Figure 13, we find that the cavity size is similar
to that of previous works within 0.07 dex. Compared to the scatter (0.17 dex), our measurements show 
a good agreement with those in the previous work. 
When we divided the cavities into two categories: cavities detected from raw images (open) and based on the $\beta$-modeling (filled), 
we do not find any significant difference in comparison. The solid vertical lines between black and red symbols show the nine overlappng cavities between \cite{Dong2010} and \cite{Panagoulia2014}. The mean difference of this subsample between two previous works is 0.06 dex, implying negligible difference. 
Note that the sample of \cite{Hlavacek2012} seems to show the largest offset (0.15 dex) among the three works,
while the difference is still smaller than the overall scatter (0.17 dex), suggesting that cavity size measurements are relatively 
consistent among various works.

\section{Summary} \label{section:summary}

Using all available X-ray data ($\sim$ 5500 exposures) from the
Chandra archive, we selected a sample of 133 targets with sufficient
X-ray photon counts to search for X-ray cavities. Based on our
consistent analysis we presented X-ray cavities detections for 69
targets over a large dynamic range from isolated galaxy to massive
galaxy clusters.  We summarize the main results as follows.\\

$\bullet$ Using various detection strategies (raw image,
$\beta$-modeling, unsharp masking method), we found 148 X-ray cavity
in 69 targets out of a total sample of 133 objects, which is $\sim$
52\% detection rate.  The detection rates of X-ray cavities are
similar among individual galaxies, galaxy groups, and galaxy clusters.
Most targets have more than two cavities, suggesting multiple AGN
outbursts. \\

$\bullet$ We confirmed the cavity size-distance relation over a large
dynamic range and environments.  We propose that the area of a
cavity is a useful parameter to represent the physical size of
cavities.\\

$\bullet$ We found little dependence of the cavity properties on
environment.  This may indicate that X-ray cavities form universally and
are not driven by global properties of the system, but
determined by the properties of AGN and its radio jets. \\

$\bullet$ The apparent relation between X-ray cavity size and X-ray
gas temperature suffers from selection effects due to the observational
limit of the spatial resolution of Chandra images as well as the
intrinsic limit of the size of X-ray emitting region, particularly for
low gas temperature systems. \\

In this work, we found that X-ray cavities have similar properties in
galaxies, groups and clusters and therefore conclude that their
formation is independent of environment. We also suggest that AGN
feedback occurs periodically based on the high fraction of X-ray
cavity detection and multiple pairs of X-ray cavities. Using our
sample, follow-up studies can investigate X-ray cavities
comprehensively as a probe of AGN feedback.

\acknowledgements This research was supported by the National Research
Foundation of Korea grant funded by the Korea government
(No. 2010-0027910 and No. 2016R1A2B3011457)
The scientific results reported in this article are based 
on data obtained from the Chandra Data Archive.  
This research has made use of software provided by the
Chandra X-ray Center (CXC) in the application packages CIAO, ChIPS,
and Sherpa.  Partial support for this work was provided by Chandra
grant AR2-13013X.


\begin{thebibliography}{106}
\expandafter\ifx\csname natexlab\endcsname\relax\def\natexlab#1{#1}\fi

\bibitem[{{Bae} \& {Woo}(2014)}]{Bae2014}
{Bae}, H.-J., \& {Woo}, J.-H. 2014, \apj, 795, 30

\bibitem[{{Barbosa} {et~al.}(2009){Barbosa}, {Storchi-Bergmann}, {Cid
  Fernandes}, {Winge}, \& {Schmitt}}]{Barbosa2009}
{Barbosa}, F.~K.~B., {Storchi-Bergmann}, T., {Cid Fernandes}, R., {Winge}, C.,
  \& {Schmitt}, H. 2009, \mnras, 396, 2

\bibitem[{{B{\^\i}rzan} {et~al.}(2008){B{\^\i}rzan}, {McNamara}, {Nulsen},
  {Carilli}, \& {Wise}}]{Birzan2008}
{B{\^\i}rzan}, L., {McNamara}, B.~R., {Nulsen}, P.~E.~J., {Carilli}, C.~L., \&
  {Wise}, M.~W. 2008, \apj, 686, 859

\bibitem[{{B{\^\i}rzan} {et~al.}(2004){B{\^\i}rzan}, {Rafferty}, {McNamara},
  {Wise}, \& {Nulsen}}]{Birzan2004}
{B{\^\i}rzan}, L., {Rafferty}, D.~A., {McNamara}, B.~R., {Wise}, M.~W., \&
  {Nulsen}, P.~E.~J. 2004, \apj, 607, 800

\bibitem[{{B{\^\i}rzan} {et~al.}(2012){B{\^\i}rzan}, {Rafferty}, {Nulsen},
  {McNamara}, {R{\"o}ttgering}, {Wise}, \& {Mittal}}]{Birzan2012}
{B{\^\i}rzan}, L., {Rafferty}, D.~A., {Nulsen}, P.~E.~J., {et~al.} 2012,
  \mnras, 427, 3468

\bibitem[{{Blanton} {et~al.}(2011){Blanton}, {Randall}, {Clarke}, {Sarazin},
  {McNamara}, {Douglass}, \& {McDonald}}]{Blanton2011}
{Blanton}, E.~L., {Randall}, S.~W., {Clarke}, T.~E., {et~al.} 2011, \apj, 737,
  99

\bibitem[{{Bogd{\'a}n} \& {Gilfanov}(2008)}]{Bogdan2008}
{Bogd{\'a}n}, {\'A}., \& {Gilfanov}, M. 2008, \mnras, 388, 56

\bibitem[{{Boroson}(2005)}]{Boroson2005}
{Boroson}, T. 2005, \aj, 130, 381

\bibitem[{{Bower} {et~al.}(2006){Bower}, {Benson}, {Malbon}, {Helly}, {Frenk},
  {Baugh}, {Cole}, \& {Lacey}}]{Bower2006}
{Bower}, R.~G., {Benson}, A.~J., {Malbon}, R., {et~al.} 2006, \mnras, 370, 645

\bibitem[{{Br{\"u}ggen} {et~al.}(2007){Br{\"u}ggen}, {Heinz}, {Roediger},
  {Ruszkowski}, \& {Simionescu}}]{Bruggen2007}
{Br{\"u}ggen}, M., {Heinz}, S., {Roediger}, E., {Ruszkowski}, M., \&
  {Simionescu}, A. 2007, \mnras, 380, L67

\bibitem[{{Br{\"u}ggen} {et~al.}(2009){Br{\"u}ggen}, {Scannapieco}, \&
  {Heinz}}]{Bruggen2009}
{Br{\"u}ggen}, M., {Scannapieco}, E., \& {Heinz}, S. 2009, \mnras, 395, 2210

\bibitem[{{Cavagnolo} {et~al.}(2010){Cavagnolo}, {McNamara}, {Nulsen},
  {Carilli}, {Jones}, \& {B{\^\i}rzan}}]{Cavagnolo2010}
{Cavagnolo}, K.~W., {McNamara}, B.~R., {Nulsen}, P.~E.~J., {et~al.} 2010, \apj,
  720, 1066

\bibitem[{{Cavaliere} \& {Fusco-Femiano}(1976)}]{Cavaliere1976}
{Cavaliere}, A., \& {Fusco-Femiano}, R. 1976, \aap, 49, 137

\bibitem[{{Chon} {et~al.}(2012){Chon}, {B{\"o}hringer}, {Krause}, \&
  {Tr{\"u}mper}}]{Chon2012}
{Chon}, G., {B{\"o}hringer}, H., {Krause}, M., \& {Tr{\"u}mper}, J. 2012, \aap,
  545, L3

\bibitem[{{Churazov} {et~al.}(2001){Churazov}, {Br{\"u}ggen}, {Kaiser},
  {B{\"o}hringer}, \& {Forman}}]{Churazov2001}
{Churazov}, E., {Br{\"u}ggen}, M., {Kaiser}, C.~R., {B{\"o}hringer}, H., \&
  {Forman}, W. 2001, \apj, 554, 261

\bibitem[{{Ciotti} {et~al.}(2010){Ciotti}, {Ostriker}, \& {Proga}}]{Ciotti2010}
{Ciotti}, L., {Ostriker}, J.~P., \& {Proga}, D. 2010, \apj, 717, 708

\bibitem[{{Clarke} {et~al.}(2004){Clarke}, {Blanton}, \&
  {Sarazin}}]{Clarke2004}
{Clarke}, T.~E., {Blanton}, E.~L., \& {Sarazin}, C.~L. 2004, \apj, 616, 178

\bibitem[{{Crenshaw} {et~al.}(2003){Crenshaw}, {Kraemer}, \&
  {George}}]{Crenshaw2003}
{Crenshaw}, D.~M., {Kraemer}, S.~B., \& {George}, I.~M. 2003, \araa, 41, 117

\bibitem[{{Croston} {et~al.}(2008){Croston}, {Hardcastle}, {Birkinshaw},
  {Worrall}, \& {Laing}}]{Croston2008}
{Croston}, J.~H., {Hardcastle}, M.~J., {Birkinshaw}, M., {Worrall}, D.~M., \&
  {Laing}, R.~A. 2008, \mnras, 386, 1709

\bibitem[{{Croton} {et~al.}(2006){Croton}, {Springel}, {White}, {De Lucia},
  {Frenk}, {Gao}, {Jenkins}, {Kauffmann}, {Navarro}, \& {Yoshida}}]{Croton2006}
{Croton}, D.~J., {Springel}, V., {White}, S.~D.~M., {et~al.} 2006, \mnras, 365,
  11

\bibitem[{{David} {et~al.}(2009){David}, {Jones}, {Forman}, {Nulsen},
  {Vrtilek}, {O'Sullivan}, {Giacintucci}, \& {Raychaudhury}}]{David2009}
{David}, L.~P., {Jones}, C., {Forman}, W., {et~al.} 2009, \apj, 705, 624

\bibitem[{{Di Matteo} {et~al.}(2005){Di Matteo}, {Springel}, \&
  {Hernquist}}]{DiMatteo2005}
{Di Matteo}, T., {Springel}, V., \& {Hernquist}, L. 2005, \nat, 433, 604

\bibitem[{{Dickey} \& {Lockman}(1990)}]{Dickey1990}
{Dickey}, J.~M., \& {Lockman}, F.~J. 1990, \araa, 28, 215

\bibitem[{{Diehl} {et~al.}(2008){Diehl}, {Li}, {Fryer}, \&
  {Rafferty}}]{Diehl2008c}
{Diehl}, S., {Li}, H., {Fryer}, C.~L., \& {Rafferty}, D. 2008, \apj, 687, 173

\bibitem[{{Dong} {et~al.}(2010){Dong}, {Rasmussen}, \& {Mulchaey}}]{Dong2010}
{Dong}, R., {Rasmussen}, J., \& {Mulchaey}, J.~S. 2010, \apj, 712, 883

\bibitem[{{Dubois} {et~al.}(2010){Dubois}, {Devriendt}, {Slyz}, \&
  {Teyssier}}]{Dubois2010}
{Dubois}, Y., {Devriendt}, J., {Slyz}, A., \& {Teyssier}, R. 2010, \mnras, 409,
  985

\bibitem[{{Dunn} \& {Fabian}(2006)}]{Dunn2006}
{Dunn}, R.~J.~H., \& {Fabian}, A.~C. 2006, \mnras, 373, 959

\bibitem[{{Dunn} {et~al.}(2005){Dunn}, {Fabian}, \& {Taylor}}]{Dunn2005}
{Dunn}, R.~J.~H., {Fabian}, A.~C., \& {Taylor}, G.~B. 2005, \mnras, 364, 1343

\bibitem[{{En{\ss}lin} \& {Heinz}(2002)}]{Ensslin2002}
{En{\ss}lin}, T.~A., \& {Heinz}, S. 2002, \aap, 384, L27

\bibitem[{{Fabian}(1994)}]{Fabian1994}
{Fabian}, A.~C. 1994, \araa, 32, 277

\bibitem[{{Fabian}(2012)}]{Fabian2012}
---. 2012, \araa, 50, 455

\bibitem[{{Fabian} {et~al.}(2002){Fabian}, {Celotti}, {Blundell}, {Kassim}, \&
  {Perley}}]{Fabian2002}
{Fabian}, A.~C., {Celotti}, A., {Blundell}, K.~M., {Kassim}, N.~E., \&
  {Perley}, R.~A. 2002, \mnras, 331, 369

\bibitem[{{Fabian} {et~al.}(2006){Fabian}, {Sanders}, {Taylor}, {Allen},
  {Crawford}, {Johnstone}, \& {Iwasawa}}]{Fabian2006}
{Fabian}, A.~C., {Sanders}, J.~S., {Taylor}, G.~B., {et~al.} 2006, \mnras, 366,
  417

\bibitem[{{Fabian} {et~al.}(2000){Fabian}, {Sanders}, {Ettori}, {Taylor},
  {Allen}, {Crawford}, {Iwasawa}, {Johnstone}, \& {Ogle}}]{Fabian2000}
{Fabian}, A.~C., {Sanders}, J.~S., {Ettori}, S., {et~al.} 2000, \mnras, 318,
  L65

\bibitem[{{Ferrarese} \& {Merritt}(2000)}]{Ferrarese2000}
{Ferrarese}, L., \& {Merritt}, D. 2000, \apjl, 539, L9

\bibitem[{{Forman} {et~al.}(2007){Forman}, {Jones}, {Churazov}, {Markevitch},
  {Nulsen}, {Vikhlinin}, {Begelman}, {B{\"o}hringer}, {Eilek}, {Heinz},
  {Kraft}, {Owen}, \& {Pahre}}]{Forman2007}
{Forman}, W., {Jones}, C., {Churazov}, E., {et~al.} 2007, \apj, 665, 1057

\bibitem[{{Gaibler} {et~al.}(2012){Gaibler}, {Khochfar}, {Krause}, \&
  {Silk}}]{Gaibler2012}
{Gaibler}, V., {Khochfar}, S., {Krause}, M., \& {Silk}, J. 2012, \mnras, 425,
  438

\bibitem[{{Ganguly} {et~al.}(2007){Ganguly}, {Brotherton}, {Cales}, {Scoggins},
  {Shang}, \& {Vestergaard}}]{Ganguly2007}
{Ganguly}, R., {Brotherton}, M.~S., {Cales}, S., {et~al.} 2007, \apj, 665, 990

\bibitem[{{Gaspari} {et~al.}(2012){Gaspari}, {Brighenti}, \&
  {Temi}}]{Gaspari2012}
{Gaspari}, M., {Brighenti}, F., \& {Temi}, P. 2012, \mnras, 424, 190

\bibitem[{{Gaspari} {et~al.}(2011){Gaspari}, {Melioli}, {Brighenti}, \&
  {D'Ercole}}]{Gaspari2011}
{Gaspari}, M., {Melioli}, C., {Brighenti}, F., \& {D'Ercole}, A. 2011, \mnras,
  411, 349

\bibitem[{{Gebhardt} {et~al.}(2000){Gebhardt}, {Bender}, {Bower}, {Dressler},
  {Faber}, {Filippenko}, {Green}, {Grillmair}, {Ho}, {Kormendy}, {Lauer},
  {Magorrian}, {Pinkney}, {Richstone}, \& {Tremaine}}]{Gebhardt2000}
{Gebhardt}, K., {Bender}, R., {Bower}, G., {et~al.} 2000, \apjl, 539, L13

\bibitem[{{Giacintucci} {et~al.}(2011){Giacintucci}, {O'Sullivan}, {Vrtilek},
  {David}, {Raychaudhury}, {Venturi}, {Athreya}, {Clarke}, {Murgia},
  {Mazzotta}, {Gitti}, {Ponman}, {Ishwara-Chandra}, {Jones}, \&
  {Forman}}]{Giacintucci2011}
{Giacintucci}, S., {O'Sullivan}, E., {Vrtilek}, J., {et~al.} 2011, \apj, 732,
  95

\bibitem[{{Gitti} {et~al.}(2012){Gitti}, {Brighenti}, \&
  {McNamara}}]{Gitti2012}
{Gitti}, M., {Brighenti}, F., \& {McNamara}, B.~R. 2012, Advances in Astronomy,
  2012

\bibitem[{{Gitti} {et~al.}(2010){Gitti}, {O'Sullivan}, {Giacintucci}, {David},
  {Vrtilek}, {Raychaudhury}, \& {Nulsen}}]{Gitti2010}
{Gitti}, M., {O'Sullivan}, E., {Giacintucci}, S., {et~al.} 2010, \apj, 714, 758

\bibitem[{{Granato} {et~al.}(2004){Granato}, {De Zotti}, {Silva}, {Bressan}, \&
  {Danese}}]{Granato2004}
{Granato}, G.~L., {De Zotti}, G., {Silva}, L., {Bressan}, A., \& {Danese}, L.
  2004, \apj, 600, 580

\bibitem[{{Guo}(2015)}]{Guo2015}
{Guo}, F. 2015, \apj, 803, 48

\bibitem[{{Heckman} \& {Best}(2014)}]{Heckman2014}
{Heckman}, T.~M., \& {Best}, P.~N. 2014, \araa, 52, 589

\bibitem[{{Hlavacek-Larrondo} {et~al.}(2012){Hlavacek-Larrondo}, {Fabian},
  {Edge}, {Ebeling}, {Sanders}, {Hogan}, \& {Taylor}}]{Hlavacek2012}
{Hlavacek-Larrondo}, J., {Fabian}, A.~C., {Edge}, A.~C., {et~al.} 2012, \mnras,
  421, 1360

\bibitem[{{Hlavacek-Larrondo} {et~al.}(2015){Hlavacek-Larrondo}, {McDonald},
  {Benson}, {Forman}, {Allen}, {Bleem}, {Ashby}, {Bocquet}, {Brodwin},
  {Dietrich}, {Jones}, {Liu}, {Reichardt}, {Saliwanchik}, {Saro}, {Schrabback},
  {Song}, {Stalder}, {Vikhlinin}, \& {Zenteno}}]{Hlavacek2015}
{Hlavacek-Larrondo}, J., {McDonald}, M., {Benson}, B.~A., {et~al.} 2015, \apj,
  805, 35

\bibitem[{{Hodges-Kluck} {et~al.}(2010{\natexlab{a}}){Hodges-Kluck},
  {Reynolds}, {Cheung}, \& {Miller}}]{Hodges2010}
{Hodges-Kluck}, E.~J., {Reynolds}, C.~S., {Cheung}, C.~C., \& {Miller}, M.~C.
  2010{\natexlab{a}}, \apj, 710, 1205

\bibitem[{{Hodges-Kluck} {et~al.}(2010{\natexlab{b}}){Hodges-Kluck},
  {Reynolds}, {Miller}, \& {Cheung}}]{Hodges2010b}
{Hodges-Kluck}, E.~J., {Reynolds}, C.~S., {Miller}, M.~C., \& {Cheung}, C.~C.
  2010{\natexlab{b}}, \apjl, 717, L37

\bibitem[{{Hopkins} {et~al.}(2006){Hopkins}, {Hernquist}, {Cox}, {Di Matteo},
  {Robertson}, \& {Springel}}]{Hopkins2006}
{Hopkins}, P.~F., {Hernquist}, L., {Cox}, T.~J., {et~al.} 2006, \apjs, 163, 1

\bibitem[{{Hudson} {et~al.}(2010){Hudson}, {Mittal}, {Reiprich}, {Nulsen},
  {Andernach}, \& {Sarazin}}]{Hudson2010}
{Hudson}, D.~S., {Mittal}, R., {Reiprich}, T.~H., {et~al.} 2010, \aap, 513, A37

\bibitem[{{Ishibashi} {et~al.}(2013){Ishibashi}, {Fabian}, \&
  {Canning}}]{Ishibashi2013}
{Ishibashi}, W., {Fabian}, A.~C., \& {Canning}, R.~E.~A. 2013, \mnras, 431,
  2350

\bibitem[{{Johnstone} {et~al.}(2002){Johnstone}, {Allen}, {Fabian}, \&
  {Sanders}}]{Johnstone2002}
{Johnstone}, R.~M., {Allen}, S.~W., {Fabian}, A.~C., \& {Sanders}, J.~S. 2002,
  \mnras, 336, 299

\bibitem[{{Kauffmann} \& {Haehnelt}(2000)}]{Kauffmann2000}
{Kauffmann}, G., \& {Haehnelt}, M. 2000, \mnras, 311, 576

\bibitem[{{Komossa} {et~al.}(2008){Komossa}, {Xu}, {Zhou}, {Storchi-Bergmann},
  \& {Binette}}]{Komossa2008}
{Komossa}, S., {Xu}, D., {Zhou}, H., {Storchi-Bergmann}, T., \& {Binette}, L.
  2008, \apj, 680, 926

\bibitem[{{Kormendy} \& {Ho}(2013)}]{Kormendy2013}
{Kormendy}, J., \& {Ho}, L.~C. 2013, \araa, 51, 511

\bibitem[{{Lal} {et~al.}(2010){Lal}, {Kraft}, {Forman}, {Hardcastle}, {Jones},
  {Nulsen}, {Evans}, {Croston}, \& {Lee}}]{Lal2010}
{Lal}, D.~V., {Kraft}, R.~P., {Forman}, W.~R., {et~al.} 2010, \apj, 722, 1735

\bibitem[{{Lanz} {et~al.}(2010){Lanz}, {Jones}, {Forman}, {Ashby}, {Kraft}, \&
  {Hickox}}]{Lanz2010}
{Lanz}, L., {Jones}, C., {Forman}, W.~R., {et~al.} 2010, \apj, 721, 1702

\bibitem[{{Li} {et~al.}(2006){Li}, {Lapenta}, {Finn}, {Li}, \&
  {Colgate}}]{Li2006}
{Li}, H., {Lapenta}, G., {Finn}, J.~M., {Li}, S., \& {Colgate}, S.~A. 2006,
  \apj, 643, 92

\bibitem[{{Li} \& {Bryan}(2014)}]{Li2014}
{Li}, Y., \& {Bryan}, G.~L. 2014, \apj, 789, 153

\bibitem[{{Machacek} {et~al.}(2010){Machacek}, {O'Sullivan}, {Randall},
  {Jones}, \& {Forman}}]{Machacek2010}
{Machacek}, M.~E., {O'Sullivan}, E., {Randall}, S.~W., {Jones}, C., \&
  {Forman}, W.~R. 2010, \apj, 711, 1316

\bibitem[{{Magorrian} {et~al.}(1998){Magorrian}, {Tremaine}, {Richstone},
  {Bender}, {Bower}, {Dressler}, {Faber}, {Gebhardt}, {Green}, {Grillmair},
  {Kormendy}, \& {Lauer}}]{Magorrian1998}
{Magorrian}, J., {Tremaine}, S., {Richstone}, D., {et~al.} 1998, \aj, 115, 2285

\bibitem[{{McNamara} \& {Nulsen}(2007)}]{McNamara2007}
{McNamara}, B.~R., \& {Nulsen}, P.~E.~J. 2007, \araa, 45, 117

\bibitem[{{McNamara} {et~al.}(2005){McNamara}, {Nulsen}, {Wise}, {Rafferty},
  {Carilli}, {Sarazin}, \& {Blanton}}]{McNamara2005}
{McNamara}, B.~R., {Nulsen}, P.~E.~J., {Wise}, M.~W., {et~al.} 2005, \nat, 433,
  45

\bibitem[{{McNamara} {et~al.}(2000){McNamara}, {Wise}, {Nulsen}, {David},
  {Sarazin}, {Bautz}, {Markevitch}, {Vikhlinin}, {Forman}, {Jones}, \&
  {Harris}}]{McNamara2000}
{McNamara}, B.~R., {Wise}, M., {Nulsen}, P.~E.~J., {et~al.} 2000, \apjl, 534,
  L135

\bibitem[{{Nakamura} {et~al.}(2006){Nakamura}, {Li}, \& {Li}}]{Nakamura2006}
{Nakamura}, M., {Li}, H., \& {Li}, S. 2006, \apj, 652, 1059

\bibitem[{{Nakamura} {et~al.}(2007){Nakamura}, {Li}, \& {Li}}]{Nakamura2007}
---. 2007, \apj, 656, 721

\bibitem[{{Nesvadba} {et~al.}(2011){Nesvadba}, {De Breuck}, {Lehnert}, {Best},
  {Binette}, \& {Proga}}]{Nesvadba2011}
{Nesvadba}, N.~P.~H., {De Breuck}, C., {Lehnert}, M.~D., {et~al.} 2011, \aap,
  525, A43

\bibitem[{{Nesvadba} {et~al.}(2008){Nesvadba}, {Lehnert}, {De Breuck},
  {Gilbert}, \& {van Breugel}}]{Nesvadba2008}
{Nesvadba}, N.~P.~H., {Lehnert}, M.~D., {De Breuck}, C., {Gilbert}, A.~M., \&
  {van Breugel}, W. 2008, \aap, 491, 407

\bibitem[{{Nulsen} {et~al.}(2009){Nulsen}, {Jones}, {Forman}, {Churazov},
  {McNamara}, {David}, \& {Murray}}]{Nulsen2009}
{Nulsen}, P., {Jones}, C., {Forman}, W., {et~al.} 2009, in American Institute
  of Physics Conference Series, Vol. 1201, American Institute of Physics
  Conference Series, ed. S.~{Heinz} \& E.~{Wilcots}, 198--201

\bibitem[{{Nulsen} {et~al.}(2005{\natexlab{a}}){Nulsen}, {Hambrick},
  {McNamara}, {Rafferty}, {Birzan}, {Wise}, \& {David}}]{Nulsen2005b}
{Nulsen}, P.~E.~J., {Hambrick}, D.~C., {McNamara}, B.~R., {et~al.}
  2005{\natexlab{a}}, \apjl, 625, L9

\bibitem[{{Nulsen} {et~al.}(2005{\natexlab{b}}){Nulsen}, {McNamara}, {Wise}, \&
  {David}}]{Nulsen2005a}
{Nulsen}, P.~E.~J., {McNamara}, B.~R., {Wise}, M.~W., \& {David}, L.~P.
  2005{\natexlab{b}}, \apj, 628, 629

\bibitem[{{O'Sullivan} {et~al.}(2011{\natexlab{a}}){O'Sullivan}, {Giacintucci},
  {David}, {Gitti}, {Vrtilek}, {Raychaudhury}, \& {Ponman}}]{OSullivan2011a}
{O'Sullivan}, E., {Giacintucci}, S., {David}, L.~P., {et~al.}
  2011{\natexlab{a}}, \apj, 735, 11

\bibitem[{{O'Sullivan} {et~al.}(2010){O'Sullivan}, {Giacintucci}, {David},
  {Vrtilek}, \& {Raychaudhury}}]{OSullivan2010}
{O'Sullivan}, E., {Giacintucci}, S., {David}, L.~P., {Vrtilek}, J.~M., \&
  {Raychaudhury}, S. 2010, \mnras, 407, 321

\bibitem[{{O'Sullivan} {et~al.}(2011{\natexlab{b}}){O'Sullivan}, {Worrall},
  {Birkinshaw}, {Trinchieri}, {Wolter}, {Zezas}, \&
  {Giacintucci}}]{OSullivan2011b}
{O'Sullivan}, E., {Worrall}, D.~M., {Birkinshaw}, M., {et~al.}
  2011{\natexlab{b}}, \mnras, 416, 2916

\bibitem[{{O'Sullivan} {et~al.}(2012){O'Sullivan}, {Giacintucci}, {Babul},
  {Raychaudhury}, {Venturi}, {Bildfell}, {Mahdavi}, {Oonk}, {Murray},
  {Hoekstra}, \& {Donahue}}]{OSullivan2012}
{O'Sullivan}, E., {Giacintucci}, S., {Babul}, A., {et~al.} 2012, \mnras, 424,
  2971

\bibitem[{{Panagoulia} {et~al.}(2014){Panagoulia}, {Fabian}, {Sanders}, \&
  {Hlavacek-Larrondo}}]{Panagoulia2014}
{Panagoulia}, E.~K., {Fabian}, A.~C., {Sanders}, J.~S., \& {Hlavacek-Larrondo},
  J. 2014, \mnras, 444, 1236

\bibitem[{{Park} {et~al.}(2012){Park}, {Kelly}, {Woo}, \& {Treu}}]{Park2012}
{Park}, D., {Kelly}, B.~C., {Woo}, J.-H., \& {Treu}, T. 2012, \apjs, 203, 6

\bibitem[{{Pounds} {et~al.}(2003){Pounds}, {Reeves}, {King}, {Page}, {O'Brien},
  \& {Turner}}]{Pounds2003}
{Pounds}, K.~A., {Reeves}, J.~N., {King}, A.~R., {et~al.} 2003, \mnras, 345,
  705

\bibitem[{{Rafferty} {et~al.}(2006){Rafferty}, {McNamara}, {Nulsen}, \&
  {Wise}}]{Rafferty2006}
{Rafferty}, D.~A., {McNamara}, B.~R., {Nulsen}, P.~E.~J., \& {Wise}, M.~W.
  2006, \apj, 652, 216

\bibitem[{{Randall} {et~al.}(2011){Randall}, {Forman}, {Giacintucci}, {Nulsen},
  {Sun}, {Jones}, {Churazov}, {David}, {Kraft}, {Donahue}, {Blanton},
  {Simionescu}, \& {Werner}}]{Randall2011}
{Randall}, S.~W., {Forman}, W.~R., {Giacintucci}, S., {et~al.} 2011, \apj, 726,
  86

\bibitem[{{Russell} {et~al.}(2010){Russell}, {Fabian}, {Sanders}, {Johnstone},
  {Blundell}, {Brandt}, \& {Crawford}}]{Russell2010}
{Russell}, H.~R., {Fabian}, A.~C., {Sanders}, J.~S., {et~al.} 2010, \mnras,
  402, 1561

\bibitem[{{Russell} {et~al.}(2013){Russell}, {McNamara}, {Edge}, {Hogan},
  {Main}, \& {Vantyghem}}]{Russell2013}
{Russell}, H.~R., {McNamara}, B.~R., {Edge}, A.~C., {et~al.} 2013, \mnras, 432,
  530

\bibitem[{{Sanders} \& {Fabian}(2002)}]{Sanders2002}
{Sanders}, J.~S., \& {Fabian}, A.~C. 2002, \mnras, 331, 273

\bibitem[{{Sazonov} {et~al.}(2006){Sazonov}, {Revnivtsev}, {Gilfanov},
  {Churazov}, \& {Sunyaev}}]{Sazonov2006}
{Sazonov}, S., {Revnivtsev}, M., {Gilfanov}, M., {Churazov}, E., \& {Sunyaev},
  R. 2006, \aap, 450, 117

\bibitem[{{Scannapieco} {et~al.}(2012){Scannapieco}, {Wadepuhl}, {Parry},
  {Navarro}, {Jenkins}, {Springel}, {Teyssier}, {Carlson}, {Couchman}, {Crain},
  {Dalla Vecchia}, {Frenk}, {Kobayashi}, {Monaco}, {Murante}, {Okamoto},
  {Quinn}, {Schaye}, {Stinson}, {Theuns}, {Wadsley}, {White}, \&
  {Woods}}]{Scannapieco2012}
{Scannapieco}, C., {Wadepuhl}, M., {Parry}, O.~H., {et~al.} 2012, \mnras, 423,
  1726

\bibitem[{{Shurkin} {et~al.}(2008){Shurkin}, {Dunn}, {Gentile}, {Taylor}, \&
  {Allen}}]{Shurkin2008}
{Shurkin}, K., {Dunn}, R.~J.~H., {Gentile}, G., {Taylor}, G.~B., \& {Allen},
  S.~W. 2008, \mnras, 383, 923

\bibitem[{{Sijacki} \& {Springel}(2006)}]{Sijacki2006}
{Sijacki}, D., \& {Springel}, V. 2006, \mnras, 366, 397

\bibitem[{{Sijacki} {et~al.}(2007){Sijacki}, {Springel}, {Di Matteo}, \&
  {Hernquist}}]{Sijacki2007}
{Sijacki}, D., {Springel}, V., {Di Matteo}, T., \& {Hernquist}, L. 2007,
  \mnras, 380, 877

\bibitem[{{Silk}(2005)}]{Silk2005}
{Silk}, J. 2005, \mnras, 364, 1337

\bibitem[{{Springel} {et~al.}(2005){Springel}, {Di Matteo}, \&
  {Hernquist}}]{Springel2005}
{Springel}, V., {Di Matteo}, T., \& {Hernquist}, L. 2005, \mnras, 361, 776

\bibitem[{{Strickland} {et~al.}(2004){Strickland}, {Heckman}, {Colbert},
  {Hoopes}, \& {Weaver}}]{Strickland2004}
{Strickland}, D.~K., {Heckman}, T.~M., {Colbert}, E.~J.~M., {Hoopes}, C.~G., \&
  {Weaver}, K.~A. 2004, \apjs, 151, 193

\bibitem[{{Strickland} {et~al.}(2000){Strickland}, {Heckman}, {Weaver}, \&
  {Dahlem}}]{Strickland2000}
{Strickland}, D.~K., {Heckman}, T.~M., {Weaver}, K.~A., \& {Dahlem}, M. 2000,
  \aj, 120, 2965

\bibitem[{{Sulentic} {et~al.}(2000){Sulentic}, {Marziani}, \&
  {Dultzin-Hacyan}}]{Sulentic2000}
{Sulentic}, J.~W., {Marziani}, P., \& {Dultzin-Hacyan}, D. 2000, \araa, 38, 521

\bibitem[{{Sutherland} \& {Dopita}(1993)}]{Sutherland1993}
{Sutherland}, R.~S., \& {Dopita}, M.~A. 1993, \apjs, 88, 253

\bibitem[{{Tombesi} {et~al.}(2010){Tombesi}, {Cappi}, {Reeves}, {Palumbo},
  {Yaqoob}, {Braito}, \& {Dadina}}]{Tombesi2010}
{Tombesi}, F., {Cappi}, M., {Reeves}, J.~N., {et~al.} 2010, \aap, 521, A57

\bibitem[{{Tremaine} {et~al.}(2002){Tremaine}, {Gebhardt}, {Bender}, {Bower},
  {Dressler}, {Faber}, {Filippenko}, {Green}, {Grillmair}, {Ho}, {Kormendy},
  {Lauer}, {Magorrian}, {Pinkney}, \& {Richstone}}]{Tremaine2002}
{Tremaine}, S., {Gebhardt}, K., {Bender}, R., {et~al.} 2002, \apj, 574, 740

\bibitem[{{Vikhlinin} {et~al.}(2006){Vikhlinin}, {Kravtsov}, {Forman}, {Jones},
  {Markevitch}, {Murray}, \& {Van Speybroeck}}]{Vikhlinin2006}
{Vikhlinin}, A., {Kravtsov}, A., {Forman}, W., {et~al.} 2006, \apj, 640, 691

\bibitem[{{Wang} {et~al.}(2011){Wang}, {Wang}, {Zhou}, {Liu}, {Wang}, {Yuan},
  \& {Dong}}]{Wang2011}
{Wang}, H., {Wang}, T., {Zhou}, H., {et~al.} 2011, \apj, 738, 85

\bibitem[{{Weymann} {et~al.}(1991){Weymann}, {Morris}, {Foltz}, \&
  {Hewett}}]{Weymann1991}
{Weymann}, R.~J., {Morris}, S.~L., {Foltz}, C.~B., \& {Hewett}, P.~C. 1991,
  \apj, 373, 23

\bibitem[{{Wise} {et~al.}(2007){Wise}, {McNamara}, {Nulsen}, {Houck}, \&
  {David}}]{Wise2007}
{Wise}, M.~W., {McNamara}, B.~R., {Nulsen}, P.~E.~J., {Houck}, J.~C., \&
  {David}, L.~P. 2007, \apj, 659, 1153

\bibitem[{{Woo} {et~al.}(2015){Woo}, {Bae}, {Son}, \& {Karouzos}}]{Woo2016}
{Woo}, J.-H., {Bae}, H.-J., {Son}, D., \& {Karouzos}, M. 2015, ArXiv e-prints

\bibitem[{{Woo} {et~al.}(2013){Woo}, {Schulze}, {Park}, {Kang}, {Kim}, \&
  {Riechers}}]{Woo2013}
{Woo}, J.-H., {Schulze}, A., {Park}, D., {et~al.} 2013, \apj, 772, 49

\bibitem[{{Zubovas} {et~al.}(2013){Zubovas}, {Nayakshin}, {King}, \&
  {Wilkinson}}]{Zubovas2013}
{Zubovas}, K., {Nayakshin}, S., {King}, A., \& {Wilkinson}, M. 2013, \mnras,
  433, 3079

\end{thebibliography}
\end{document}